%% file: main.tex
\documentclass[11pt]{report}
\usepackage[utf8]{inputenc}
\usepackage{times}
\usepackage{geometry}
\usepackage{enumitem,amssymb}
\usepackage{wrapfig,lipsum,booktabs}
\usepackage{ragged2e}
\usepackage{amsmath}
\usepackage{graphicx}
\usepackage{comment}
\usepackage{eurosym}
\usepackage{multicol}
\usepackage[usenames]{xcolor} 
\definecolor{xlinkcolor}{cmyk}{1,1,0,0}
\usepackage{url}
\usepackage[
 colorlinks=true,    
 linkcolor=xlinkcolor,     
 citecolor=xlinkcolor,     
 filecolor=xlinkcolor,  
 urlcolor=xlinkcolor,      
 final=true
]{hyperref}
\usepackage[super,sort&compress]{natbib}
\usepackage{enumitem}
\setenumerate{itemsep=0mm}
\usepackage{authblk}
\usepackage[belowskip=-5pt,aboveskip=-5pt]{caption}
\usepackage{subfiles}
\usepackage{tocbibind}
\usepackage[symbol]{footmisc}

\usepackage{color, colortbl}

\title{\vspace{-2cm}
Theories and Experiments for Testable Baryogenesis Mechanisms\\ {\small A Snowmass White Paper}
}
\author[1]{J. L. Barrow\thanks{jbarrow@fnal.gov; also at Tel Aviv University}}
\affil[1]{The Massachusetts Institute of Technology, Department of Physics, Cambridge, MA 02139, USA}
\author[2]{Leah Broussard}
\affil[2]{Oak Ridge National Laboratory}
\author[3]{James M.\ Cline}
\affil[3]{McGill University, Montréal, Canada}
\author[4]{P. S. Bhupal Dev}
\affil[4]{Washington University in St. Louis}
\author[5]{Marco Drewes}
\affil[5]{Universit\'e catholique de Louvain, Louvain-la-Neuve B-1348, Belgium}
\author[6]{Gilly Elor}
\affil[6]{PRISMA$^+$ Cluster of Excellence \& Mainz Institute for Theoretical Physics\\
Johannes Gutenberg University, 55099 Mainz, Germany}
\author[7]{Susan Gardner}
\affil[7]{University of Kentucky, Lexington}
\author[8]{Jacopo Ghiglieri}
\affil[8]{SUBATECH,  Nantes Universit\'e , IMT Atlantique, IN2P3/CNRS, Nantes, France}
\author[9]{Julia Harz}
\affil[9]{Physik Department T70, Technische Universität München,
James-Franck-Straße 1, 85748 Garching, Germany}
\author[10]{Yuri Kamyshkov}
\affil[10]{The University of Tennessee at Knoxville}
\author[5]{Juraj Klaric}
\author[11]{Lisa W. Koerner}
\affil[11]{University of Houston}
\author[3]{Benoit Laurent}
\author[12]{Robert McGehee}
\affil[12]{University of Michigan, Ann Arbor}
\author[13]{Marieke Postma}
\affil[13]{Nikhef, Amsterdam, Netherlands}
\author[14]{Bibhushan Shakya}
\affil[14]{Deutsches Elektronen-Synchrotron DESY, Notkestr. 85, 22607 Hamburg, Germany}
\author[15]{Robert Shrock}
\affil[15]{Stony Brook University}
\author[14]{Jorinde van de Vis}
\author[16]{Graham White\thanks{graham.white@ipmu.jp}}
\affil[16]{The University of Tokyo, Kashiwa, Japan}
\date{\today}
\begin{document}

\maketitle


\noindent { \bf Abstract:} The baryon asymmetry of the Universe is one of the central motivations to expect physics beyond the Standard Model. In this Snowmass white paper, we review the challenges and opportunities in testing some of the central paradigms that predict physics at scales low enough to expect new experimental data in the next decade. Focusing on theoretical ideas and some of their experimental implications, in particular, we discuss neutron-antineutron transformations, flavor observables, next generation colliders, future neutron facilities, gravitational waves, searches for permanent electric dipole moments, $0\nu \beta \beta $ decay and some future large underground experiments as methods to test post-sphaleron baryogenesis, electroweak baryogenesis, mesogenesis and low scale leptogenesis. Finally, we comment on the cases where high scale physics can be probed through some of these same mechanisms.

\newcommand{\mychapter}[2]{
    \setcounter{chapter}{#1}
    \setcounter{section}{0}
    \chapter*{#2}
    \addcontentsline{toc}{chapter}{#2}
}

\makeatletter
\newcommand{\chapterauthor}[1]{%
  {\parindent0pt\vspace*{-7pt}%
  \linespread{1.1}\normalsize\centering\scshape#1%
  \par\nobreak\vspace*{5pt}}
  \@afterheading%
}
\makeatother

\mychapter{0}{Executive Summary}
\subfile{subfiles/ExecutiveSummary}

\tableofcontents
\markboth{}{}






\mychapter{1}{Post-sphaleron Baryogenesis}
\subfile{subfiles/PSB}

\subfile{subfiles/Probes}

\mychapter{2}{Electroweak Baryogenesis}
\subfile{subfiles/EWBG}

\mychapter{3}{Mesogenesis}
\subfile{subfiles/Mesogenesis}

\mychapter{4}{Leptogenesis}

\subfile{subfiles/lowscalelepto}
\newpage
\bibliographystyle{style}    
\bibliography{bibliography}   

\end{document}

%% file: subfiles/ExecutiveSummary.tex
The origins of the baryon asymmetry of the Universe is a fundamental mystery 
and is a key motivation for physics beyond the Standard Model. 
The observed matter-antimatter asymmetry in the universe apparently obligates the laws of physics to include some mechanism of baryon number ($\mathcal{B}$) violation (BNV), as specified by the Sakharov conditions \citep{Sakharov:1967dj}. However, perturbatively, the Standard Model (SM) ``accidentally'' conserves $\mathcal{B}$; in non-perturbative regimes at temperatures of $\gtrsim10\,$TeV, electroweak sphalerons instead conserve ``good" baryon minus lepton quantum numbers ($\mathcal{B}-\mathcal{L}$) while violating $\mathcal{B+L}$\citep{tHooft:1976snw}, implying a form of BNV. Even still, as the primordial plasma moves through the electroweak phase transition, it is known that electroweak instantons (sphalerons) tend to ``wash out" any preexisting baryon asymmetry without some other form of (typically) \textit{higher-scale} lepton asymmetry production--perhaps near $10^{12}\,$GeV, as is the case in classic leptogenesis\citep{Dolgov:1991fr}. Due to the effective impossibility of definitively testing for classic leptogenesis in any kind of ``on shell" manner similar to the historic experimental confirmation of the $W^{\pm}$, $Z^0$, and Higgs, one must also begin to investigate other potentially observable baryogenesis alternatives. 
$\mathcal{B-L}$-violating $|\Delta\mathcal{B}|=2$ and $|\Delta\mathcal{L}|=2$ processes and $CP$ violation measurements are thus of foremost importance in uncovering the nature of baryogenesis. Of course, other methods and tests can prove extremely powerful as the community prioritizes the discovery of physics beyond the Standard Model.

In this vein, here we review the challenges and opportunities of four of the most promising testable baryogenesis paradigms, taking a predominately theoretical view while detailing some experimental opportunities. Post-sphaleron baryogenesis is likely to be testable through neutron-antineutron oscillations, and experiments are expected to improve current sensitivity by three orders of magnitude. Electroweak baryogenesis can be probed via high sensitivity searches for electric dipole moments, and there are prospects for compelling tests of this paradigm from colliders and gravitational waves. However, there are some unresolved theoretical issues which makes the status of this class of models unclear unless conditions are very unfavourable. The CP violation asymmetry of Standard Model mesons is an observable that is directly related to the baryon asymmetry in mechanisms of Mesogenesis.  Mesogenesis also predicts new seemingly baryon violating hadron decay modes; designated searches are already underway at Belle, BaBar and LHCb. Additional exotic decays can be searched for at future and present kaon and hyperon factories. Finally, the introduction of heavy neutral leptons with near degenerate masses can consistently and simultaneously explain both the origin of neutrino masses and the baryon asymmetry. This can be tested through beam dump experiments, colliders, meson decays, $0\nu \beta \beta$ decays, lepton flavor decays, and can also be indirectly probed through cosmological observables. Together, these four paradigms present rich opportunities for discovery which can explain the lack of antimatter in the Universe, but each will require dedicated theoretical and experimental resources and support in order to make progress toward discovery.

%% file: subfiles/PSB.tex
Post-sphaleron baryogenesis (PSB)\citep{Babu:2006xc} is an attractive low-scale mechanism able to explain the observed matter-antimatter asymmetry of the Universe. As the name suggests, the generation of the baryon asymmetry occurs after the electroweak sphalerons have gone out of equilibrium around temperatures $T\sim 130$ GeV. The same $|\Delta \mathcal{B}|=2$ operator which gives rise to baryogenesis in this scenario also leads to $n\rightarrow\bar{n}$ oscillations, thus making an intimate connection between the two $\mathcal{B}$-violating observables. In fact, it has been argued by some (in particular, by one of the proponents of electroweak baryogenesis\citep{NelsonINT2017}) that the PSB mechanism is more compelling than other baryogenesis mechanisms such as electroweak baryogenesis, because (i) it is consistent with a wide range of cosmology/inflation models; (ii) no high-temperature physics is required, which solves a lot of cosmological issues, e.g. gravitino overproduction; and (iii) electroweak baryogenesis requires a first-order phase transition and CP violation in Higgs sector which are very constrained by electric dipole moment of electron, mass of Higgs, and other LHC constraints. 

The basic idea of PSB is that there can be  baryon-number-violating decays of a heavy (pseudo)scalar $S$. It is known that $|\Delta \mathcal{B}|=1$ operators are strongly constrained by experimental proton decay limits and cannot lead to successful PSB. Thus, one can instead rely on dimension $d=9$, $|\Delta \mathcal{B}|=2$ operators of the form $qqqqqq/\Lambda^5$, which lead to $S\to 6q$ and $6\bar{q}$ decays. If these decays violate CP and occur out-of-equilibrium, then it is possible to directly generate an asymmetry between baryons and antibaryons irrespective of electroweak sphalerons, and hence, the PSB mechanism can happen even below the sphaleron freeze-out. 

\section{Prediction of an Upper Limit on $\tau_{n\rightarrow\bar{n}}$ and Recent Theoretical Progress}
\begin{figure}[htp]
    \centering
    \includegraphics[width=0.6\columnwidth]{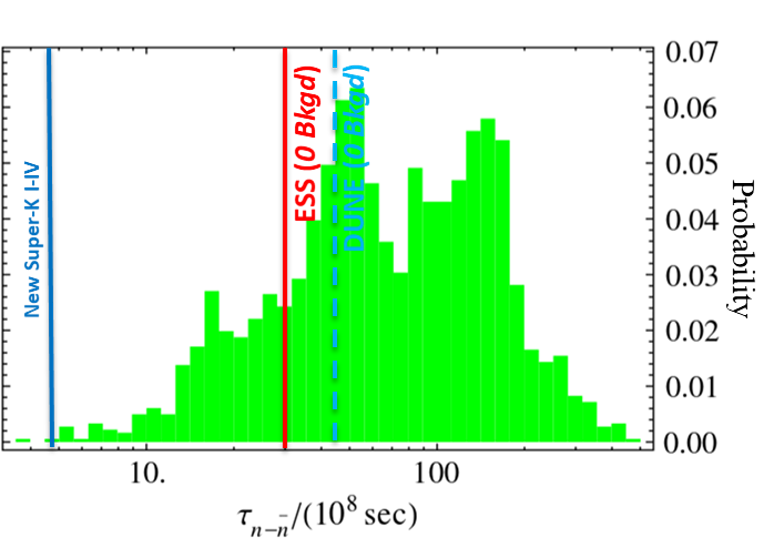}
    \caption{The probability distribution of the $n\rightarrow\bar{n}$ lifetime prediction in the PSB model\citep{Babu:2013yca}. This figure is a reproduction of Fig. 9 in work by Babu, Dev and colleagues\citep{Babu:2013yca} with overlayed sensitivity estimates for DUNE, assuming a 25\% efficiency, and ESS NNBAR, assuming an ILL-like efficiency\citep{BaldoCeolin:1994jz}. Each of these assumes a case of zero background when compared to the Super-Kamiokande I-IV limit\citep{Super-Kamiokande:2020bov}.}
    \label{fig:expectedlimits}
\end{figure}

The PSB mechanism, when embedded in a quark-lepton unified model based on the Pati-Salam gauge group $SU(4)_c\times SU(2)_L\times SU(2)_R$\citep{Pati:1974yy}, leads to an absolute upper limit on the neutron-antineutron oscillation time $\tau_{n\bar{n}}\lesssim 10^{10-11}$ sec\citep{Babu:2008rq, Babu:2013yca}, which might be within reach of future experiments such as NNBAR at the ESS and DUNE; see Figure~\ref{fig:expectedlimits}. The upper bound on the $n\rightarrow\bar{n}$ oscillation time ($\tau_{n\bar{n}}$) in this model is a consequence of the bounded range of the parameters which the amplitudes corresponding to $\tau_{n\bar{n}}$ depend upon. Requiring the model to reproduce the observed neutrino masses and other meson oscillation parameters fixes the form of the Yukawa coupling matrices involved in the generation of baryon asymmetry in this model. The given form of the matrix essentially puts an upper bound on the maximum baryon asymmetry $(\epsilon_{B}^{\rm max})$ that can be produced. This asymmetry is produced by the decay of a color-sextet real scalar $S$ to quarks and antiquarks, mediated by scalar diquarks $\Delta_{dd}$ and $\Delta_{ud}$\citep{Babu:2013yca}. This process happens at a specific decay temperature ($T_d$), depending on the mass of the scalar, $M_S$, and of the diquarks, $M_{\Delta_{ud}}$ and $M_{\Delta_{dd}}$. The observed baryon asymmetry $\eta_B$ at the recombination epoch is related to the baryon asymmetry produced in the decay of $S$ through a dilution factor proportional to $T_d/M_S$. Since the dilution cannot be more than $\eta_B/\epsilon_{B}^{\rm max}$, this sets an upper bound on the range of $M_S$. On the other hand, the $S$ must decouple while being relativistic before its decay produces the asymmetry, which requires the $\Delta_{ud}$ and $\Delta_{dd}$ masses to be at least a factor of 5-10 larger than $M_S$. If $\Delta_{ud}$ and $\Delta_{dd}$ are too heavy, however, this will drive the $T_d$ lower, and hence, result in a larger dilution. This delicate interplay between the model parameters makes this scenario quite predictive and testable. 

Since the original calculation\citep{Babu:2013yca}, there have been two important updates, namely, (i) recent lattice QCD results on the relevant $\Delta\mathcal{B}$ operator\citep{Rinaldi:2018osy, Rinaldi:2019thf}, which is 16\% smaller than the old MIT bag model estimate\citep{Rao:1982gt}, and (ii) an improved calculation of the baryon asymmetry, which takes into account both wave-function and vertex correction diagrams, along with a more accurate treatment of the color combinatorics\citep{BhupalPresentation,Proceedings:2020nzz}. The effects of these updates on the probability plot shown in Figure~\ref{fig:expectedlimits} are still being analyzed. 

\section{Other Unique Signals of PSB}
Apart from $n\rightarrow\bar{n}$, the PSB model also features multi-TeV-scale scalar diquarks, which can  be searched for in the form of high-mass dijet resonances at the LHC and other future hadron colliders\citep{Mohapatra:2007af,Chen:2008hh,Berger:2010fy,Baldes:2011mh,Chivukula:2015zma, CMS:2019gwf, Pascual-Dias:2020hxo}. In addition, if the diquarks couple to both left- and right-handed quark bilinears, they could lead to an observable neutron electric dipole moment\citep{Bell:2018mgg}. 

Another class of simplified PSB models are based on the SM gauge group, but by adding renormalizable terms which violate baryon number\citep{Dev:2015uca, Allahverdi:2017edd}. In this case, gauge invariance requires introduction of new colored fields. A minimal setup is to add iso-singlet, color-triplet scalars $X_\alpha$ with hypercharge $Y=+4/3$ to allow for terms such as $X_\alpha d^cd^c$ in the Lagrangian. We need at least two $X$'s (i.e. $\alpha=1,2$) to produce baryon asymmetry from $X$ decay. However, following the general arguments of Kolb and Wolfram\citep{Kolb:1979qa}, it can be shown that with just the $Xd^cd^c$ interaction term, the net baryon asymmetry vanishes after summing over all flavors of $d^c$. Therefore, we need additional $B$-violating interactions. A simple scenario is to introduce a SM-singlet Majorana fermion $\psi$ which can also play the role of dark matter, if its mass is close to the proton mass\citep{Allahverdi:2013mza, Allahverdi:2017edd}. After integrating out the $X$ fields, this model gives an effective $B$-violating operator $\psi u^cd^cd^c$, which also induces $n\to \bar{n}$ oscillation for Majorana $\psi$ at one-loop level. There is a nice interplay between baryon asymmetry, dark matter-baryonic matter coincidence, $n\to \bar{n}$ oscillation, as well as monojet and dijet collider signals in this model\citep{Allahverdi:2017edd}.

\section{Outlook: Implications of $n\rightarrow\bar{n}$ searches for other baryogenesis scenarios}

Generally, $n\rightarrow\bar{n}$ searches are capable of probing several baryogenesis frameworks across a broad range of energy scales. For instance, baryogenesis can be realized with decays of fermions instead of scalars or pseudoscalars as in the PSB scenario. Such instances are naturally realized in supersymmetric theories, where late decays of a gaugino (the superpartner of a gauge boson or a hidden sector counterpart), which tends to be the lightest supersymmetric particle, through $R$-parity violating couplings and in turn can produce the baryon asymmetry\citep{Cui:2012jh,Cui:2013bta,Arcadi:2015ffa,Pierce:2019ozl}. If the RPV operators involve first generation quarks, $n\rightarrow\bar{n}$ can be mediated by squarks and gauginos and is often the most promising experimental probe of such baryogenesis mechanisms.

Different EFT studies investigated the consequences of $n\rightarrow\bar{n}$ oscillations on models of high-scale baryogenesis\citep{Grojean:2018fus,Fridell:2021gag}.
It was demonstrated that the observation of $n\rightarrow\bar{n}$ oscillations at experiments such as DUNE or NNBAR would indicate that the washout arising from the effective $n-\bar{n}$ oscillation operators would be very efficient down to around 100 TeV\citep{Fridell:2021gag}. This could point towards a baryogenesis mechanism below 100 TeV, motivating to search for new physics at a future 100 TeV collider. However, for a conclusive statement, washout in all flavours has to be confirmed.

Generally, one can distinguish two topologies at tree-level which UV-complete the effective dimension-9 $n\--\bar{n}$ operators: (1) a topology with two bosons and a Majorana fermion~\citep{Zwirner:1984is,Barbieri:1985ty,Mohapatra:1986bd,Lazarides:1986jt,Goity:1994dq,Babu:2001qr,Babu:2006wz,Allahverdi:2010im,Gu:2011ff,Gu:2011fp,Allahverdi:2013mza,Dev:2015uca,Dhuria:2015swa,Ghalsasi:2015mxa,Gu:2016ghu,Calibbi:2016ukt,Gu:2017cgp,Calibbi:2017rab,Allahverdi:2017edd} and (2) a topology with a trilinear boson coupling~\citep{Mohapatra:1980qe,Mohapatra:1982xz,Chang:1984qr,Babu:2006xc,Babu:2008rq,Baldes:2011mh,Babu:2012vc,Arnold:2012sd,Babu:2013yca,Patra:2014goa,Herrmann:2014fha}. 

Ref.\citep{Grojean:2018fus} focused on a scenario along the lines of topology (1). Its EFT study of baryogenesis from fermion decays suggests that $n\rightarrow\bar{n}$ searches can reliably probe various classes of such scenarios with decaying fermion masses up to $1000\,$TeV, far beyond the reach of any conceivable high energy collider.

 Topology (2) was studied in Ref.\citep{Fridell:2021gag}, introducing two new scalar diquarks. Hereby, the heavier diquark can undergo CP-violating decays, generating the baryon asymmetry. In case $n-\bar{n}$ oscillations were observed at future experiments and if both diquarks would feature similar masses, even for maximal CP violation the EFT results are recovered~\citep{Fridell:2021gag} and the washout would be too strong to allow for successful baryogenesis. This is in contrast to the case when the new particles feature a large mass hierarchy, e.g. one near the GUT scale and the other within future collider reach. In this case $n-\bar{n}$ oscillation experiments and future collider searches can provide complementary probes. Assuming the maximal washout based on a single-flavour analysis, such a high-scale baryogenesis model remains a viable scenario to account for the observed baryon asymmetry.


%% file: subfiles/Probes.tex


\section{Neutron-Antineutron Transformations as Probes of PSB}
The search for $n\rightarrow\bar{n}$ transformations provides an important probe of low-scale baryon-number violation (BNV) that is relevant to post-sphaleron baryogenesis. Naive dimensional analysis led to an early view that proton (and bound neutron) decay would be the most sensitive test of BNV. The reasoning for this was that the operators which mediate proton decay in the low-energy effective field theory (EFT) are four-fermion operators, with coefficients having dimensions of $1/M^2$ (where $M$ was presumably the GUT mass). In contrast, the operators which mediate $n\rightarrow\bar{n}$ are six-quark operators with coefficients having dimensions of $1/M^5$. Hence, though naively, if one assumes a single high mass scale $M=\Lambda$ which is responsible for BNV, then the $n\rightarrow\bar{n}$ process would be more suppressed than proton decay. However, there exist models where proton decay is negligibly small while $n\rightarrow\bar{n}$ constitutes the primary manifestation of baryon number violation\citep{Mohapatra:1980qe,Nussinov:2001rb,Girmohanta:2019fsx,Arnold:2012sd}. Consequently, there is a strong motivation to experimentally search for $n\rightarrow\bar{n}$. 

The current best limit on $n\rightarrow\bar{n}$ comes from a free-neutron experiment performed using neutrons from a high-power reactor at the Institut Laue-Langevin (ILL). This limit is usually quoted as a lower limit on the free neutron oscillation time, $\tau_{n\rightarrow\bar{n}}$, defined by the transition probability for an initial neutron to transform to an $\bar{n}$, namely $P(n(t) \to \bar{n}) = \sin^2(t/\tau_{n\rightarrow\bar{n}})e^{-t/\tau_n}$, where $\tau_n$ denotes the mean life of the neutron; the ILL experiment\citep{BaldoCeolin:1994jz} obtained the lower limit $\tau_{n\rightarrow\bar{n}} > 0.86 \times 10^8\,$s. A new and more sensitive $n\rightarrow\bar{n}$ search experiment is currently planned at the European Spallation Source (ESS) being built and commissioned in Lund, Sweden \citep{Phillips:2014fgb,Addazi:2020nlz}. For any free $n\rightarrow\bar{n}$ experiment, one must maximize the figure of merit, $\langle N t^2 \rangle$, encapsulating the need for the maximum number of $n$s on target over the tenure of any experiment, $N$, and observing those $n$s on their respective flight paths for a maximum amount of time, $t$\cite{Phillips:2014fgb,Addazi:2020nlz}. 

The existence of free $n\rightarrow\bar{n}$ similarly implies the possibility of $\bar{n}$ annihilation within nuclei through $\{\bar{n}n,\bar{n}p\}\rightarrow N\pi$. Similarly, it generally gives rise to dinucleon decays involving ordinary nucleons within nuclei, i.e. of the form $\{nn,np,pp\}\rightarrow N\pi$. Each of these yields multi-pion final states which are rather semi-spherical in topology, a so-called ``pion star". The signature is complicated by the fact that the pions propagate through the nuclear medium, reinteracting during their propagation (also known as ``final state interactions"). A recent search for matter instability due to $n\rightarrow\bar{n}$ was performed within oxygen nuclei using the Super-Kamiokande (SK) water Cherenkov detector. From the total neutron exposure, expected backgrounds and observed candidates, a lower limit on the intranuclear lifetime $\tau_{M}$ can be constructed using Bayesian method. From this, one extracts a limit on the free lifetime $\tau_{n\rightarrow\bar{n}}$ via the relation $\tau_{n\rightarrow\bar{n}} = (\tau_{M}/R)^{1/2}$, where $R \simeq 10^{23}$ s$^{-1}$ takes account of nuclear effects (an ``intranuclear suppression factor"). The SK experiment\citep{Super-Kamiokande:2020bov} obtained a lower limit of $\tau_{n\rightarrow\bar{n}} > 4.7 \times 10^8\,$s.

Thinking toward future intranuclear searches, in order to optimize the sensitivity, one must maximize the neutron exposure. The probability of a free conversion is expected to grow as $t^2$ in the quasi-field-free limit\citep{Phillips:2014fgb,BITTER1985461,Davis:2016uyk}; in principle, this is similarly possible for conversions within the nucleus, though phenomenologically it derives from the depth of the nuclear potential well. Together, these allow a given experiment to uncover the minimum $n\rightarrow\bar{n}$ oscillation period, $\tau_{n\bar{n}}$. For the sake of time, we discuss here only the future of free neutron searches throughout the rest of this document.

\subsection{A Future for Free Neutron Conversion Experiments}
By maximizing flux, angular acceptance, background rejection, flight path length, and utilizing modern detector and reconstruction technologies, the proposed NNBAR program\citep{Phillips:2014fgb,Addazi:2020nlz} at the European Spallation Source (ESS) is projected to be able to culminate in an ultimate sensitivity increase for $n\rightarrow \bar{n}$ of three orders of magnitude ($\gtrsim1000\times$)\citep{Klinkby:2014cma,Santoro:2020nke} over that previously attained with free $n$s after the above-mentioned search at ILL, which achieved $\langle Nt^2 \rangle=1.5\times10^9n$s${}^2$ and 
the lower limit $\tau_{n\bar{n}}> 0.86\times 10^8\,$s \citep{BaldoCeolin:1994jz}. 
Taken together, the HIBEAM/NNBAR program\citep{Addazi:2020nlz} will enable the characterisation of a mixing sector involving $n$s, $\bar{n}$s, and potentially sterile neutrons, $n'$. 

The ESS, located in Lund, Sweden, is a multi-disciplinary international laboratory which will operate the world's most powerful pulsed $n$ source. The development of the facility has been driven by the $n$ scattering community, and the first 15 instruments are currently under construction. The start of the User Program is expected to begin in 2023, and the 15 current instruments represent only a subset of the full 22-instrument-suite required to fully realize the ESS' scientific mission as defined in the ESS statutes.

During the preceding era, $n$ facilities' contributions to particle physics have been limited to only a handful of experimental endeavors. In light of this, regarding later-built ESS instruments 16-22, an ESS-lead analysis of the facility’s scientific diversity has identified the need of a fundamental physics beamline, ANNI\citep{Soldner:2018ycf}/HIBEAM\citep{Addazi:2020nlz}, as \textit{the highest priority}. Beginning in the early-mid 2020's, and without the necessity of utilizing the expected full beam power of $5\,$MW available $\gtrsim2030$, the HIBEAM\citep{Addazi:2020nlz} program could utilize the fundamental physics beamline to consider dark sector-oriented searches through sterile $n'$ conversions \citep{Mohapatra:1980qe,Berezhiani:2005hv,Babu:2006xc,Dev:2015uca,Allahverdi:2017edd,Grojean:2018fus,Foot:2004pa,Foot:2014mia,Berezhiani:2020nzn} using magnetic field controls to permit $n$ disappearance ($n\rightarrow n'$) and $n$ regeneration ($n\rightarrow n' \rightarrow n$), as well as cobaryogenesis-oriented searches via $n\rightarrow n' \rightarrow\bar{n}$, and possibly a small-scale $n\rightarrow \bar{n}$ search. At full power, ANNI\citep{Soldner:2018ycf} can be utilized for other beyond Standard Model physics searches such as those for $n$ electric dipole moments.

\subsection{NNBAR at the ESS Large Beam Port}
In addition to the ANNI\citep{Soldner:2018ycf}/HIBEAM\citep{Addazi:2020nlz} fundamental physics beamline, another remarkable opportunity for the particle physics community is offered by the already constructed Large Beam Port (LBM), which in fact lies within the ESS monolith, a critical provision created specifically for the NNBAR experiment. A normal ESS beamport would be far too small for NNBAR to reach its ambitious sensitivity goals. Therefore, part of the beam extraction system in the ESS monolith has been engineered to enable the construction of a large frame covering the size of \textit{three} standard beamports. Initially, the frame will be filled by three modularized regular-size beamports which can later be removed to provide NNBAR full access to the LBP for the duration of the experiment
. It cannot be understated how such a configuration is entirely unique among currently operating $n$ facilities. The monolith interface, supporting an opening up to 1\,m$^2$, provides a substantial view of the voluminous cryogenic moderator with a time-averaged brilliance rivaling those of modern research reactors. 

\subsection{HighNESS in Support of NNBAR}
To further develop the plans for this future upgrade to ESS capabilities, and to greatly empower NNBAR, a design study\citep{Santoro:2020nke} has begun fashioning a new lower neutron moderator consisting of liquid deuterium and able to provide a high-flux of cold $n$s to benefit both NNBAR and future $n$ scattering experiments. This project, termed HighNESS, is funded by the Research and Innovation Action within the EU's Horizon 2020 program for \euro3M over the next three years. The evaluation letter from the European Commission unambiguously highlighted the importance of “the potential for discoveries in the new Physics beyond the standard model” through the NNBAR experiment. 
A key deliverable of the HighNESS project is the Conceptual Design Report of NNBAR, of which the Collaboration's recent white paper\citep{Addazi:2020nlz} is an important first step along the journey toward baryon number violation measurements.

\subsection{Technological Developments in Support of Neutron Conversion Searches}
Achieving maximum $\langle Nt^2 \rangle$ at NNBAR will utilize several key improvements to $n$ scattering and high-energy particle detectors made over the last 30 years since the original ILL experiment\citep{BaldoCeolin:1994jz}. Beyond advancements in detector materials, timing, and software reconstruction capabilities related to identifying\citep{Hewes:2017xtr} topologically spherical ``$\pi$-star" $\bar{n}$ annihilation events\citep{Golubeva:1997fs,Golubeva:2018mrz,Barrow:2019viz} on a thin ${}^{12}C$ foil surrounded by detector elements, of critical importance is the growing capabilities of advanced, high $m$-valued\citep{mezei1976novel,Granada:2020ksc,Jamalipour:2020lsp,Abele:2005xd} $n$ reflectors to guide $n$s downrange into an intense, well constrained focus. An effective reflector must be fully illuminated by the source; thus, a substantial amount of the overall $n$ beam intensity will have trajectories which deviate significantly from the nominal beam trajectory axis. To achieve this, reflecting angles for even fairly cold $n$s ($\sim$1000 m/s) will exceed that of the limit of the best traditional reflectors. NNBAR will thus utilize high $m$-valued $n$ \textit{supermirror} technologies~\cite{mezei1976novel} which have been successful as a means to guide thermal and cold $n$s to many scattering instruments at both pulsed and continuous $n$ sources. NNBAR will utilize the same multi-layered surface coating treatment on its reflector to gather and focus the wide range of $n$ trajectories incident from the source. To do so, NNBAR requires reflectors with a surface reflectivity of $m\gtrsim6$, i.e. a reflection capability six times higher than that of polished nickel. 

\subsection{NNBAR Experimental Description}
Neutrons emerging from the LBP will be reflected via a pseudoellipsoidal, differential supermirror along a magnetically shielded volume towards a ${}^{12}C$ target foil, surrounded by an annihilation detector; a beam trap downrange absorbs the beam. The detection efficiency of an annihilation event on the foil is assumed to be $\gtrsim50\%$ as in the ILL experiment\citep{BaldoCeolin:1994jz}; work is underway to assess whether such detection is also expected to be backgroundless\citep{Santoro:2020nke}, as before\citep{BaldoCeolin:1994jz}. 

Assuming an experimental duration of three years, these improvements collectively provide an enhancement in the $\langle Nt^2 \rangle$ figure of merit by $\gtrsim 1000\times$ILL using \textit{only} a lower moderator, enabling an increase in $\tau_{n\bar{n}}$ by $\gtrsim 30\times$ILL. 
A full quantification of the NNBAR sensitivity is part of the HighNESS program. However, it should be noted that, in principle, running times can be extended to mitigate against any loss of sensitivity. Furthermore, estimates provided in\citep{Addazi:2020nlz,FrostPhD} are rather conservative with an assumed selection efficiency for an annihilation event of $\sim 50$\% as obtained at the ILL; indeed, detector technology and data analysis methods in experimental particle physics are now substantially more advanced, and so a far higher efficiency could be expected. These, together with the opportunity to view \textit{both} the upper and lower moderators, provide ample ability to mitigate any unexpected losses in sensitivity, and can empower an unprecedented discovery. 


%% file: subfiles/EWBG.tex
\section{Introduction}
In the early Universe, temperatures are expected to be high enough for electroweak symmetry to be restored. At such a high temperature, non perturbative processes that violate $B+L$, known as sphalerons,\citep{Manton:1983nd,Klinkhamer:1984di} are very efficient. If electroweak symmetry is eventually broken during a strongly first order phase transition, bubbles of broken electroweak vacuum form and expand until they fill the entire universe. The sphaleron rate goes out of equilibrium inside these bubbles. If there are CP violating interactions with the bubble wall, a chiral asymmetry accumulates outside the bubble wall which biases the sphalerons into producing a net baryon asymmetry. Some of this asymmetry is then swept up by the advancing bubble wall, in which the net baryon number remains conserved as the sphaleron rate becomes negligible.
\par
\subsection{New states at the electroweak scale} \par  
Although the Standard Model contains all the qualitative ingredients for electroweak baryogenesis to be catalyzed, quantitatively it fails on two counts. First the ratio of the W boson to Higgs mass is far too small to catalyze a first order transition.\citep{Bochkarev:1987wf, Kajantie:1995kf} The electroweak transition is expected instead to be a smooth crossover. Second, the CP violation that exists in the CKM matrix cannot catalyze enough of a chiral asymmetry to seed the observed baryon asymmetry.\citep{Gavela:1993ts, Huet:1994jb, Gavela:1994dt} We therefore require new states to modify the Standard Model predictions. 

\subsection{Models} \par 
Early research on electroweak baryogenesis focused on the MSSM.\citep{Cohen:1992zx,Carena:1996wj,Riotto:1997vy,Cline:1997vk,Cline:2000kb,Cline:2000nw,Lee:2004we,Chung:2009qs,Cirigliano:2006wh} This particular model is likely dead\citep{Liebler:2015ddv} although the final nail in the coffin is elusive. Other early work focused on the 2HDM\citep{Turok:1990zg,Dine:1990fj,Fromme:2006cm,Liu:2011jh} which requires near non-perturbative couplings to achieve a strong first order phase transition.\citep{Kainulainen:2019kyp} The two Higgs doublet model faces strong experimental constraints from negative results in looking for electric dipole moments, though there is still some viable parameter space.\citep{Bian:2014zka, Fuyuto:2019svr} 
The severe constraints on the MSSM and 2HDM has led the community to consider other models that are less constrained by electric dipole moments. The degree of constraint from EDMs qualitatively depends upon unresolved theoretical issues (forthcoming). Viable models that have recently been considered include extending the standard model by a single vector-like fermion,\citep{Egana-Ugrinovic:2017jib} the standard model effective theory with CP violation in the $\tau$ lepton sector,\citep{DeVries:2018aul} singlet extensions of the Standard Model augmented by CP violating operators,\citep{Cline:2012hg,Vaskonen:2016yiu} singlet extensions of the MSSM (the NMSSM) \citep{Cheung:2012pg,Bian:2017wfv,Akula:2017yfr} and composite Higgs Models.\citep{Bruggisser:2018mrt}
The community has also recently considered models that avoid strong constraints on CP violation. For example, the amount of CP violation needed is less if the weak sphaleron rate is faster than expected in the early Universe due to dynamical gauge coupling constants.\citep{Ellis:2019flb} The CP violation can also be hidden in a dark sector.\citep{Cline:2017qpe,Carena:2018cjh} Finally, electroweak symmetry breaking could have occurred at much higher temperatures \citep{Baldes:2018nel,Glioti:2018roy} or in a multistep process where sphalerons first go out of equilibrium at TeV temperatures,\citep{Inoue:2015pza} possibly having some zero temperature symmetries spontaneously broken at high temperature.\citep{Ramsey-Musolf:2017tgh}

\subsection{Electroweak baryogenesis searches at colliders}
Since new particles at the weak scale are needed to make the electroweak phase transition (EWPT) strongly first order, and to provide sufficient CP violation, EWBG models have strong potential for discovery at the Large Hadron Collider (LHC) or future colliders.  A popular means of strengthening the EWPT is by coupling the Higgs to a singlet scalar $s$,
which can enhance the cubic term in the Higgs thermal potential \citep{Anderson:1991zb}
or provide a tree-level potential barrier in a two-step phase transition where $s$ gets a vacuum expectation value.\citep{Espinosa:2011ax} LHC constraints on invisible Higgs decays strongly constrain the case where the mass of the singlet is smaller than half of the Higgs mass (especially when the singlet is stable).\citep{Cline:2013gha} \ \   In the regime where the scalars are heavier than half a Higgs mass, they are typically difficult to observe in colliders,\citep{Curtin:2014jma} although singlet mixing could be detected at a Higgs factory.\citep{Huang:2016cjm} Projections show that the HL-LHC will be able to constrain deviations in the triple Higgs interaction, which are a generic feature of extended scalar sectors, up to 50\% at 68\% confidence level.\citep{Cepeda:2019klc} \ \   If the singlet also participates in the new CP-violating interactions, LHC discovery prospects can be strong.\citep{Cline:2021iff}\ \   The enhanced CP-violating interactions require new states with masses $\lesssim 1$\,TeV that are constrained by collider searches.  These include leptophilic $Z'$ gauge bosons,
\citep{Carena:2018cjh}
stau-like particles, 
\citep{Cline:2017qpe} vectorlike top partners \citep{Cline:2021iff}. Additionaly, the CP-violating interactions can affect Higgs branching ratios, \citep{Chien:2015xha, Bahl:2020wee} and induce flavor-changing neutral currents. 

\subsection{Experimental searches for permanent electric dipole moments} \par
In some models for EWBG, the CP-violating interactions at zero temperature are many orders of magnitude larger than CP-violating from the CKM-matrix and can be measured in EDM experiments. CP-violating operators contribute to the (chromo)-EDMs via two-loop Barr-Zee diagrams and to the 2-loop Weinberg three gluon operator, see e.g.\citep{Engel:2013lsa} for a review. The bound on the electron-EDM from the ACME experiment, $d_e < 1.1 \times 10^{-29} e\, {\rm cm}$ at 90\% C.L.\citep{ACME:2018yjb}, has ruled out many baryogenesis models. The most stringent bound on the neutron EDM, which gets contributions from the quark EDMs, quark chromo-EDMs and the Weinberg three gluon operator, is $d_n<1.8 \times 10^{-26} e\, {\rm cm}$ at 90\% C.L. \citep{Abel:2020pzs}. The interpretation of this bound as a constraint on CP-violating operators is challenged by the uncertainty of hadronic and nuclear matrix elements.\citep{Chien:2015xha} Further improvement of the sensitivity of EDM experiments by about an order of magnitude is expected in the coming years for the neutron EDM\citep{PhysRevC.97.012501, nEDM:2019qgk, Wurm:2019yfj, n2EDM:2021yah} as well as the electron EDM.\citep{2013NJPh...15e3034T,PhysRevLett.119.133002,2018EPJD...72..197A}


\subsection{Theoretical issues with CP violating sources}\par 
To compute the baryon asymmetry, one must solve a network of coupled Boltzmann equations for CP-asymmetric perturbations  
to distributions of particle species in the plasma.  Historically the field has been divided between two methodologies for computing the CP-violating source term for these equations, known as the VEV-insertion approximation (VIA) \citep{Riotto:1997vy,Riotto:1995hh} and semiclassical or WKB approach,
\citep{Joyce:1994zt,Cline:2000nw} respectively.  Although the source terms from the competing formalisms have some qualitative similarities, the VIA source term systematically leads to estimates for the baryon asymmetry
that can be orders of magnitude larger than the WKB predictions,\citep{Cline:2020jre} especially in theories where the CP-violating interactions reside in light fermions.\citep{Cline:2021dkf} The viability of EWBG models that can still be probed in the future by EDM experiments relies critically on the existence of the VIA source term. \ \

The consistency of the VIA has been questioned.  For example it is known \citep{Lee:2004we} that the VIA source term is often infrared divergent if thermal widths $\Gamma_T$ of particles are neglected, which is difficult to justify physically.  The convergence of the VIA expansion requires the bubble wall thickness to exceed  $\Gamma_T$,\citep{Postma:2021zux} which is unrealistic in many models.  A recent reconsideration of the VIA source term derivation from first principles concludes that it vanishes in the simplest models \citep{Kainulainen:2021oqs}.  These issues are still under vigorous investigation, and one may hope for some convergence within the community in the near future.

\subsection{Bubble wall velocity} \par 
The determination of the wall velocity is an important theoretical issue, as a larger baryon asymmetry is generally produced by slower walls. More precisely, the baryon asymmetry is expected to go to zero smoothly as $v_w\rightarrow1$.\citep{Cline:2020jre, Dorsch:2021ubz}\ \  In principle, the wall velocity can be computed from the conservation of the energy-momentum tensor and a set of Boltzmann equations,\citep{Moore:1995si} but this procedure has not yet been carried out in a fully consistent way. The main issue arises from the Boltzmann equation for the background species, which is singular at $v_w=c_s=1/\sqrt{3}$ and causes the friction on the wall to diverge at this velocity. It is still unclear what causes this singularity, as some studies claim that it is an artifact of truncating moments of the Boltzmann equation,\citep{Laurent:2020gpg} while another suggests that it is a consequence of the linearization of the Boltzmann equation.\citep{Dorsch:2021nje}\ \  Nevertheless, significant progress has recently been made to characterize the importance of equilibrium hydrodynamic effects \citep{Konstandin:2010dm,BarrosoMancha:2020fay,Balaji:2020yrx,Cline:2021iff,Ai:2021kak} for avoiding the divergence of the out-of-equilibrium perturbations. An important conclusion for baryogenesis is that hydrodynamic effects heat up the plasma around the wall when it approaches the speed of sound, increasing the pressure on the wall, which is maximized at the Jouguet velocity. This implies that most walls can be classified in two distinct groups:\citep{Cline:2021iff} weak phase transitions  are stopped by the pressure peak and become deflagration walls with a velocity close to the speed of sound $0.5\lesssim v_w\lesssim 0.6$, and phase transition strong enough to overcome the pressure peak become detonation walls with an ultrarelativistic velocity $\gamma_w\gg 1$. 

\subsection{Other theoretical challenges}
Beyond the major issues of calculating the CPV source and the bubble wall velocity, there are some other theoretical issues that require attention. First, it is unclear how to best model the deviation from equilibrium, as most studies use a specific ansatz, but others\citep{Cline:2020jre,Laurent:2020gpg} suggest that it can cause unphysical singularities of the linearized Boltzmann equation, and opt instead for a truncation scheme of the moment expansion. Second, in approaches to the Boltzmann equations involving more than just a single moment of the distribution functions, new collision terms involving the sphaleron rates are required, which have not been computed in lattice calculations and hence are difficult to quantify.\citep{Dorsch:2021ubz}\ \ Third, it is unclear if there is a strong model dependence in the baryon washout condition which has only been calculated for a handful of models.\citep{Ahriche:2007jp,Funakubo:2009eg,Fuyuto:2014yia, Fuyuto:2015jha} The thermal widths and rates are typically computed at leading order and need updating. Finally, at finite temperature perturbation theory converges slowly even after performing daisy resummation.\citep{Croon:2020cgk,Gould:2021oba}\ \ A more accurate determination of the thermal potential could have large effects on the strength of the phase transition, as well as the speed and shape of the bubble wall.

\subsection{Potential signals in Gravitational waves} \par
When the bubbles of broken electroweak vacuum collide, they typically source a gravitational wave signal. Most relevant for electroweak baryogenesis is the scenario where the bubble wall does not reach an ultra-relativistic velocity. The main contribution to the gravitational wave signal then comes from the sound waves in the plasma, and the subsequent turbulence regime.\citep{Caprini:2019egz} It was demonstrated\citep{Cline:2020jre, Dorsch:2021ubz} that successful baryogenesis can in principle occur for wall velocities that are large enough to source a stochastic gravitational wave signal that can be observed with LISA. However, in minimal models\citep{Cline:2021iff, Lewicki:2021pgr}, there is only a limited overlap between the parameters preferred for successful baryogenesis and the parameters associated to a gravitational wave signal observable by LISA. Next-generation space-based gravitational wave experiments such as DECIGO\citep{Kawamura:2020pcg} and BBO\citep{Harry:2006fi} will have an increased sensitivity, and will thus be able to probe a much larger part of parameter space.

%% file: subfiles/Mesogenesis.tex
\section{Introduction}
\emph{A contribution by G. Elor and R. McGehee} \\
Mesogenesis is a new, experimentally testable mechanism of MeV-scale baryogenesis and dark matter production which leverages the CPV in SM meson systems~\cite{Elor:2018twp,Elor:2020tkc,Elahi:2021jia}. 
Generic to all ``flavors" of Mesogenesis is a scalar field $\Phi$ with a mass of $10-100 \text{ GeV}$ which decays at a low temperature $T_R$ to $q \bar{q}$ pairs. $\Phi$ may or may not be related to inflation, but $T_{\rm BBN} \lesssim T_R \lesssim T_{\rm QCD}$, 
so there exists a late matter-dominated era. As the temperature is below the QCD phase transition, the $q \bar{q}$'s subsequently hadronize into SM neutral and charged mesons which undergo out-of-equilibrium CPV processes such as neutral $B^0_{d,s}$ oscillations or charged meson decays. These processes are expected in the SM, but CPV contributions from new physics could exist. Baryon number is never violated thanks to the introduction of a new dark sector fermion $\psi_{\mathcal{B}}$ carrying baryon number $B=-1$. There are two sub-classes of Mesogenesis models which we discuss in detail below.

\subsection{Dark Baryons}

\begin{figure}[h!]
    \centering
    \includegraphics[width=\textwidth]{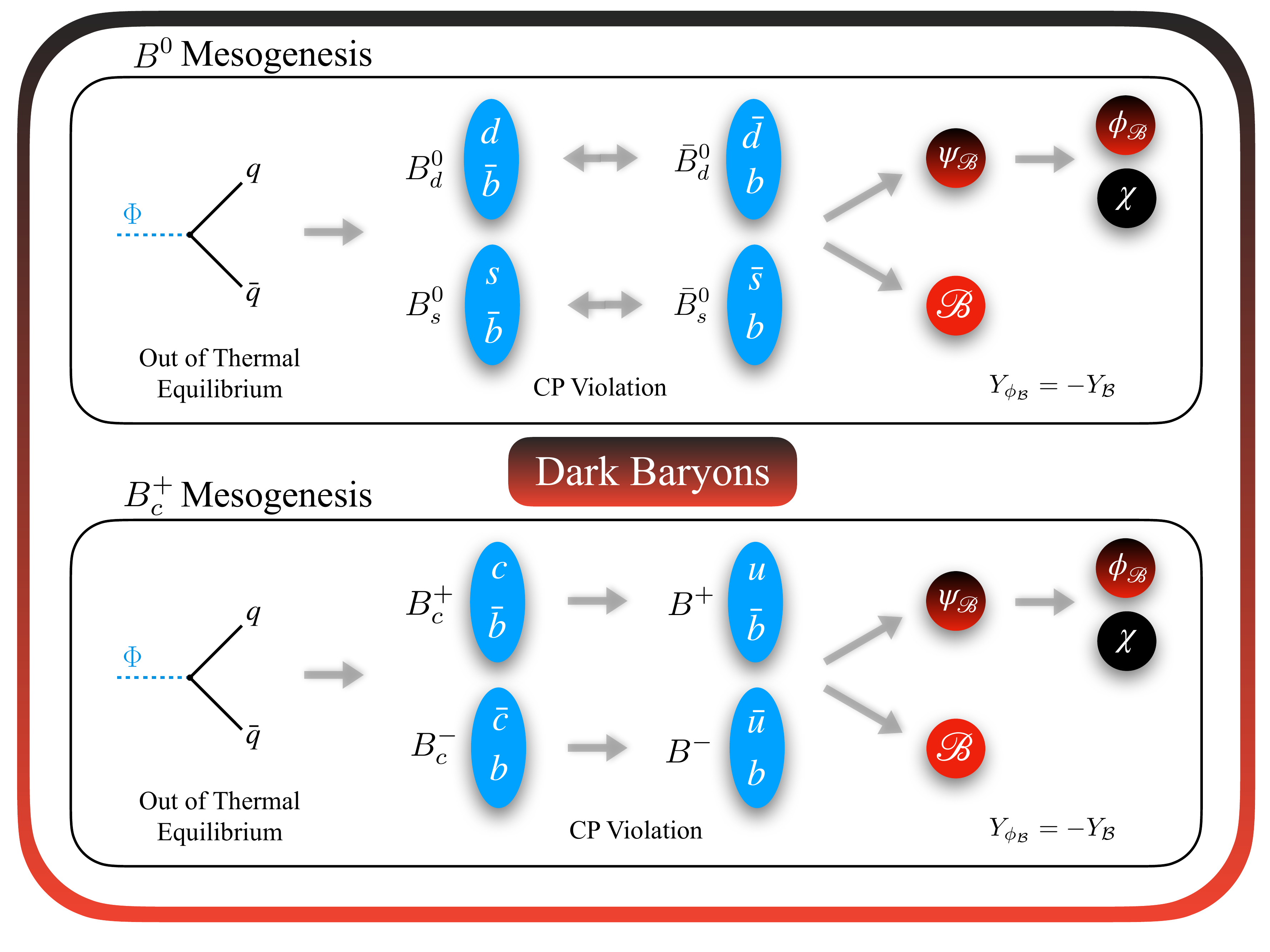}
    \caption{Illustration of how  Neutral $B$ Mesogenesis (top) and $B_c^+$ Mesogenesis (bottom) satisfy the Sakhorov conditions generating an equal and opposite baryon asymmetry in the dark and visible sectors. }
    \label{fig:MesoDarkBaryon}
\end{figure}

First, we consider a scenario in which the daughter meson of the CPV process decays into the dark baryon and a SM baryon, the result of which is the generation of an equal and opposite baryon asymmetry between the dark and visible sectors. The stability of matter requires $m_{\mathcal{B}} \gtrsim m_p$, so this is only possible for sufficiently heavy daughter mesons. 
This mass choice does permit hyperon decays to dark baryon final states and searches at present and future hyperon factories could yield additional  constraints~\cite{Alonso-Alvarez:2021oaj, Goudzovski:2022vbt}.

\subsubsection{Neutral $B$ Mesogenesis}
In Neutral $B$ Mesogenesis \cite{Elor:2018twp}, the CPV of $B_{s,d}^0-\bar{B}_{s,d}^0$ oscillations is leveraged. A colored triplet scalar $Y$ with electric charge $-1/3$ and baryon number $-2/3$ is introduced to mediate the decay into the dark baryon. The following interactions are allowed $\mathcal{L}_Y = - \sum_{i,j} y_{ij} Y^* \bar{u}_{i,R} d_{j,R}^c - \sum_k y_{\psi_{\mathcal{B}}k} Y \bar{\psi}_{\mathcal{B}} d_{kR}^c  + \text{h.c.}$. 
Consistency with LHC bounds requires $M_Y \sim \mathcal{O}(\rm TeV)$, so integrating out this scalar yields the effective operator: 
\begin{equation}
    \label{eq:EffOp}
    \mathcal{O}_{b u_i d_j} = \frac{y^2}{M_Y^2} \bar{u}_i^c d_j \bar{b}^c \psi_{\mathcal{B}} + \text{h.c.}
\end{equation}
where $y^2 \equiv y_{ij} y_{\psi_{\mathcal{B}} 3}$. This allows the $b$ quark within the neutral $B$ mesons to decay via $\bar{b}\rightarrow \psi_{\mathcal{B}} u d$. After undergoing oscillations, $B$ mesons decay into dark and SM baryons, resulting in an equal and opposite baryon asymmetry between the dark and visible sectors. The baryon asymmetry is directly linked to experimental observables and successful Mesogenesis requires
\begin{equation}
    A_{\rm sl} \times \text{Br}\left( B^0 \rightarrow \mathcal{B}_{\rm SM} + \psi_{\mathcal{B}} \right) \gtrsim 10^{-7} \,,
\end{equation}
where $A_{\rm sl}$ is the semi-leptonic asymmetry. 
Searches for the apparent baryon-number-violating meson decays in this mechanism are already underway at Belle~\cite{Belle:2021gmc} and LHCb \cite{Rodriguez:2021urv,Borsato:2021aum}. Furthermore, the UV model giving rise to Eq.~\eqref{eq:EffOp} also predicts new decay modes for hyperons baryons \cite{Alonso-Alvarez:2021oaj,Goudzovski:2022vbt} which are being searched for by BES \cite{BESIII:2021slv}. Since neutral $B$ Mesogenesis predicts the existence of a colored triplet mediator, collider and flavor observables can indirectly probe this mechanism. Given the plethora of signals and ongoing experimental searches, neutral $B$ Mesogenesis is likely to be fully probed within the next 5-10 years \cite{Alonso-Alvarez:2021qfd}. Neutral $B$ Mesogenesis can also be explicitly realized in a supersymmetric model with Dirac Gauginos and an $R$-symmetry identified with baryon number \cite{Alonso-Alvarez:2019fym}.

\begin{figure}[t!]
    \centering
    \includegraphics[width=\textwidth]{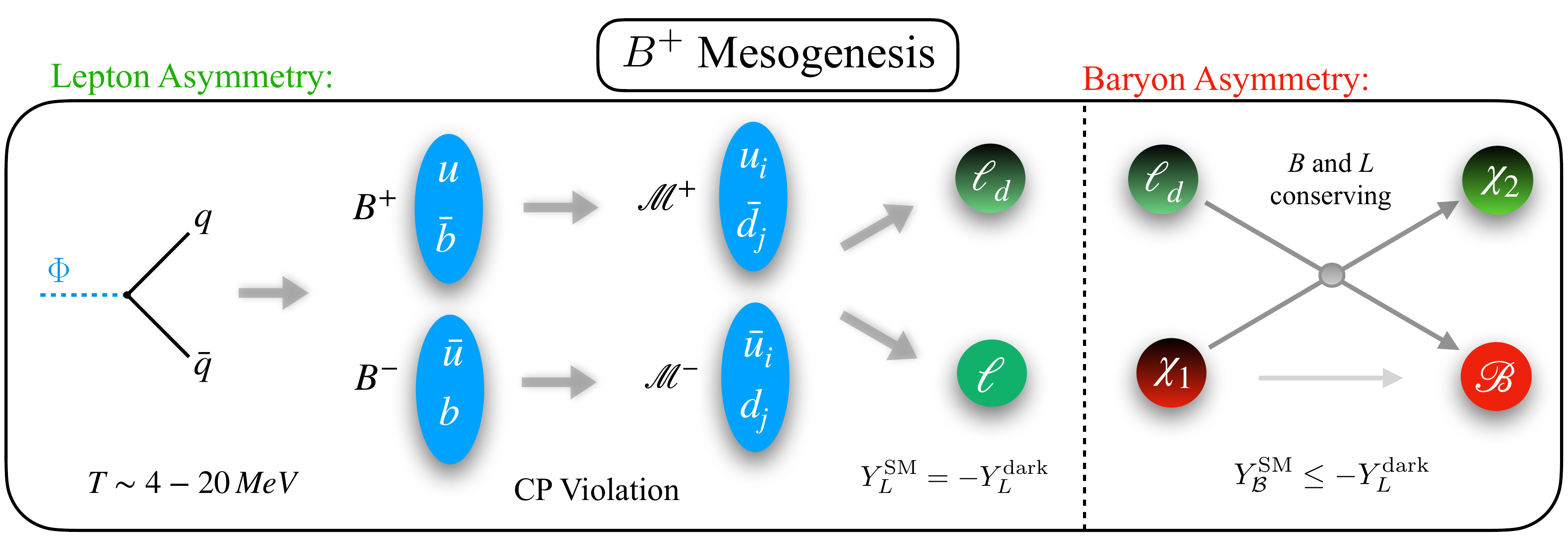}
    \vspace{0.5cm}\caption{Depiction of how $B^+$ Mesogenesis satisfies the Sakhorov conditions. $D^+$ Mesogenesis proceeds the same way, with the initial charged $B$s replaced by $D$s.}
    \label{fig:MesoDarkLepton}
\end{figure}
 
\subsubsection{$B_c^+$ Mesogenesis}
In $B_c^+$ Mesogenesis \cite{Elahi:2021jia}, $B_c^+$ mesons undergo CPV decays to $B^+$ mesons which subsequently decay into the dark sector via the operator in Eq.~\eqref{eq:EffOp}: 
\begin{subequations}\label{eq:BcMech}
\begin{align}
B_c^+ \to \, & B^+ + f \,, \, \quad  B^+ \to \, \psi_{\mathcal{B}} + \mathcal{B}^+.
\end{align}
\end{subequations}
The baryon asymmetry is directly controlled by 1) the CPV in $B_c^+$ decays, 2) the branching fraction of the $B_c^+$ decay into $B^+$ mesons and other SM final states, and 3) the branching fraction of the $B^+$ meson into SM baryons and missing energy. The first observable is expected to be sizeable~\cite{Choi:2009ym} but is currently not well constrained, nor is the second. However, the branching fraction of $B^+$ is being probed by the same searches as neutral $B$ Mesogenesis. 
Overall, this is a remarkably simple model of Mesogenesis and provides motivations for $B_c$ physics searches at e.g. LHCb~\cite{Gouz:2002kk} and the Future Circular Collider~\cite{FCC:2018byv}. 

\subsection{Dark Leptons}


In the second class of Mesogenesis models, the daughter mesons of the CPV process are too light to decay to a pair of dark and SM baryons. Instead, they decay into a pair of dark and SM leptons resulting in an equal and opposite lepton asymmetry between the dark and visible sectors. This lepton asymmetry is then transferred to a baryon asymmetry between the two sectors via dark-sector processes. 

Two such models of Mesogenesis involve CPV decays of $D^+$~\cite{Elor:2020tkc} and $B^+$~\cite{Elahi:2021jia} mesons:
\begin{subequations}
\begin{align}
 D^+ \, \text{or} \, B^+ \, \to \, &\mathcal{M}^+ \,+\,  \mathcal{M} \,, \quad
\mathcal{M}^+  \, \to \, \ell_d \,+\, \ell^+ \,, 
\end{align}
\end{subequations}
where $\mathcal{M}^+$ is a charged SM meson. Since these decays occur at $\mathcal{O}(10 \text{ MeV})$ temperatures, electroweak sphalerons cannot convert the SM lepton asymmetry into a baryon asymmetry. Instead, dark-sector scattering is used to efficiently transfer the lepton asymmetry to a baryon asymmetry. 

The generated lepton asymmetry is directly tied to experimental observables such as the CPV in a particular decay mode:
\begin{subequations}
\begin{align}
A_{\rm CP} = \frac{\Gamma(D^+ \rightarrow f) - \Gamma(D^- \rightarrow \bar{f})}{\Gamma(D^+ \rightarrow f) + \Gamma(D^- \rightarrow \bar{f})}
\end{align}
\end{subequations}
(with an analogous definition for $B^+$ decays). To achieve a lepton asymmetry greater than the observed baryon asymmetry, the relevant CPV and branching ratios in each Mesogenesis model must satisfy
\begin{subequations}
\begin{align}
D^+: \quad \sum_{f \supset \pi^+} A_{\rm CP}^f \text{Br}(D^+ \to f) \gtrsim 3\times 10^{-5}, \quad \text{Br}(\pi^+ \to \ell_d + \ell^+) \gtrsim 10^{-3},\\
B^+: \quad \sum_{f \supset \mathcal{M}^+} A_{\rm CP}^f \text{Br} (B^+ \to f) \gtrsim 5.4 \times 10^{-5}, \quad \sum_{\mathcal{M}^+} \text{Br} (\mathcal{M}^+ \to \ell_d + \ell^+) \gtrsim 10^{-3}.
\end{align}
\end{subequations}
$D^+$ Mesogenesis may thus be probed by improved sensitivity to both CPV and branching ratios of $D^+$ decays to pions (at \emph{e.g.} LHCb) and $\text{Br}(\pi^+ \to \ell_d + \ell^+)$. 
For $B^+$ Mesogenesis, it is possible that the SM contains the necessary CPV and branching ratios required to produce the observed baryon asymmetry. Although it is difficult to calculate $A_{\rm CP}^f$, some predicted branching fractions of $B^+$ are on the order of the current experimental central values~\cite{Beneke:2005vv}. It is instead easier to probe the decays of the lighter $\mathcal{M}^+$ to SM leptons $+$ invisible (\emph{e.g.}~\cite{Hayano:1982wu,E949:2014gsn,PIENU:2017wbj,NA62:2017qcd,PIENU:2019usb,NA62:2021bji}), often by recasting searches for sterile neutrinos.

\subsection{Testability}

\begin{figure}[t!]
    \centering
    \includegraphics[width=\textwidth]{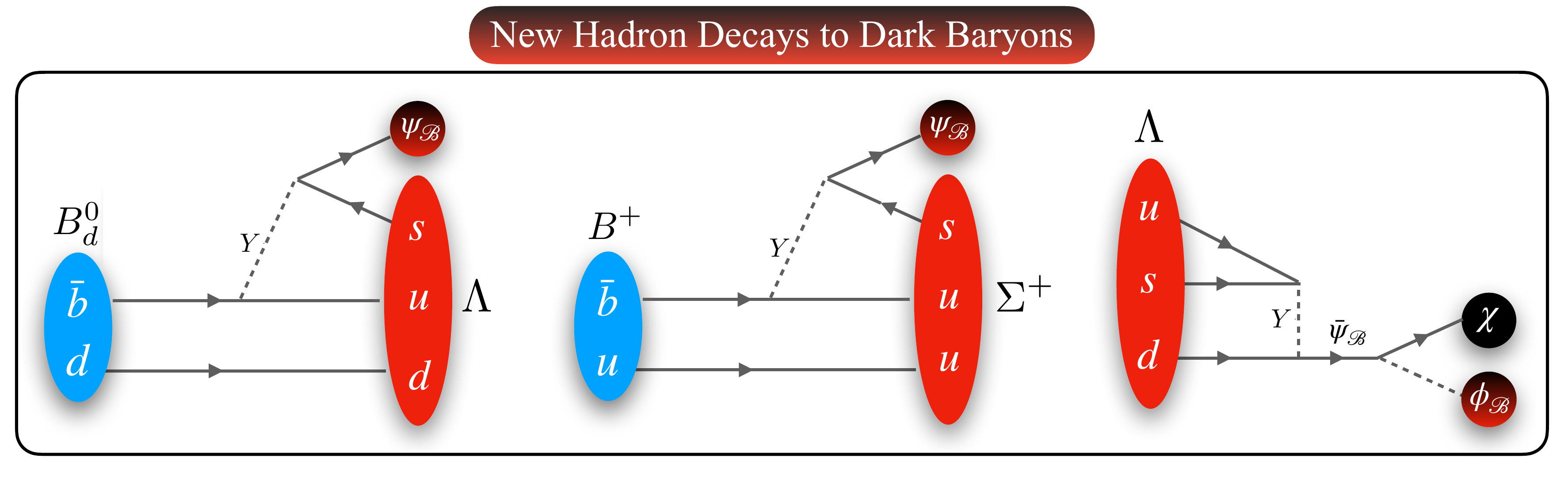}
    \caption{Some examples of apparently baryon-number-violating decay modes of SM hadrons into dark baryons. In the left and middle diagrams, the $B$ meson decay proceeds through $\mathcal{O}_{bus}$. The left decay $B_d^0 \rightarrow \Lambda^0 \psi_{\mathcal{B}}$ is directly related to the baryon asymmetry produced in the neutral $B$ Mesogenesis mechanism. The central diagram $B^+ \rightarrow \Sigma^+ \psi_{\mathcal{B}}$ is an observable directly related to $B_c^+$ Mesogenesis. In the first, a designated search targeting Mesogenesis by the Belle Collaboration \cite{Belle:2021gmc} has probed this operator. In the right-most diagram, the fully invisible decay of the $\Lambda$ baryon indirectly probes neutral $B$ and $B_c^+$ Mesogenesis. A recent search by BESIII \cite{BESIII:2021slv} has targeted exactly this decay mode. See \cite{Alonso-Alvarez:2021oaj} for other $s$-flavored baryon decays arising in these models.}
    \label{fig:DarkBaryonDecays}
\end{figure}

Table.~\ref{tab:decayChannels} summarizes the four Mesogenesis mechanisms discussed above. In all cases, the baryon asymmetry is directly related to experimental observables that can be probed by various colliders. For each \emph{flavor} of Mesogenesis, we itemize the relevant observables along with the experiments most suited to probe them in Table.~\ref{tab:decayChannels}. 
\begin{table*}[h]
\renewcommand{\arraystretch}{1}
  \setlength{\arrayrulewidth}{.25mm}
\centering
\small
\setlength{\tabcolsep}{0.2 em}
\begin{tabular}{ | c | c | c | c | c  |  c|}
    \hline
  Mechanism & CPV & Dark Sector & Observables  &  Relevant Experiments \\
    \hline \hline
    $B^0$  Mesogenesis 
    &  $B_s^0 \,\, \& \,\, B_d^0$   
    &  dark baryons   
    & $A^{s,d}_{sl}$    
    & LHCb     \\  \cite{Elor:2018twp} 
    &   oscillations
    &  
    & $\text{Br} (B\rightarrow \mathcal{B}+ X)$
    &  $B$ Factories, LHCb    \\
    \hline
    
    &  
    &   
    & $A_{CP}^D$  
    & $B$ Factories, LHCb \\
    $D^+$  Mesogenesis  
    &  $D^\pm$ decays   
    &   dark leptons  
    & $\text{Br}_{D^+}$  
    &  $B$ Factories, LHCb \\ \cite{Elor:2020tkc}
    &    
    &    and baryons
    &  $\text{Br} (\mathcal{M^+} \rightarrow \ell^+ + X)$  
    &  peak searches e.g. PSI, PIENU \\
    \hline
     & 
     & 
     & $A_{CP}^B$
     & $B$ Factories, LHCb\\
    $B^+$  Mesogenesis  
     &  $B^\pm$ decays   
     & dark leptons 
     & $\text{Br}_{B^+}$
     &$B$ Factories, LHCb   \\ \cite{Elahi:2021jia} 
     &   
     &  and baryons
     &  $\text{Br} (\mathcal{M^+} \rightarrow \ell^+ + X)$  
     & peak searches e.g. PSI, PIENU  \\
    \hline
     &  
     &
     & $A_{CP}^{B_c}$ 
     &  LHCb, FCC \\
     $B^+_c$  Mesogenesis 
     &  $B^\pm_c$ decays & dark baryons 
     & $\text{Br}_{B_c^+}$ 
     & LHCb, FCC \\ \cite{Elahi:2021jia}
     &  
     &
     &  $\text{Br}_{B^+\rightarrow \mathcal{B}^++ X}$ 
     &  $B$ Factories, LHCb \\
    \hline

\end{tabular}\vspace{0.5cm}
\caption{Overview of the four Mesogenesis mechanisms, their experimental observables, and the most relevant experiments to probe them.
}
\label{tab:decayChannels}
\end{table*}
Both neutral $B$ Mesogenesis and $B_c^+$ Mesogenesis require the introduction of a TeV scalar mediator and a dark baryon which allows SM mesons and baryons to undergo apparent baryon-number-violating decays. Such new decay modes are either directly related to the generated baryon asymmetry or serve as indirect probes of the mechanisms; we highlight a few example decay modes in Fig.~\ref{fig:DarkBaryonDecays}. 

In $B^+$ and $D^+$ Mesogenesis, the lepton asymmetry is related to apparently lepton-number-violating meson decays. Since the lepton asymmetry is then transferred to the baryon asymmetry via dark sector scatterings, few model-independent probes exist. However, a UV embedding would give rise to additional signals related to the dark sector. 


%% file: subfiles/lowscalelepto.tex
The experimental observation of neutrino ($\nu$) masses indicates the existence of new elementary particles beyond the SM.
At tree level it can be conveniently
explained by tree-level physics through the \emph{see-saw} 
mechanisms of
type-I\citep{Minkowski:1977sc,Mohapatra:1979ia,Glashow:1979nm,GellMann:1980vs,Yanagida:1980xy,Schechter:1980gr} 
  type-II\citep{Schechter:1980gr,Magg:1980ut,Cheng:1980qt,Lazarides:1980nt,Mohapatra:1980yp} and type-III\citep{Foot:1988aq}. In addition, many radiative models exist\citep{Cai:2017jrq}, including many that can connect to the Dark Matter problem\citep{Restrepo:2013aga}. 
Leptogenesis\citep{Fukugita:1986hr} is an attractive scenario in which the matter-antimatter asymmetry of the observable universe\citep{Canetti:2012zc} is related to the properties of $\nu$s, allowing to constrain the parameter space with data from $\nu$ oscillation experients.
While leptogenesis was first proposed in the type I seesaw, it can also be realised with scalar\citep{Ma:1998dx} or fermionic\citep{Hambye:2003rt,Albright:2003xb} triplets or combinations thereof\citep{Antusch:2004xy}, cf.\citep{Hambye:2012fh} for a review, but the type I incarnation remains to be the most studied one on which we focus here.

\section{Low scale seesaw and leptogenesis}

A generic prediction of the type I seesaw is the existence of Heavy Neutral Leptons (HNLs) with Majorana masses near the seesaw scale.
The Yukawa couplings of the HNLs to the lepton doublets introduce
new sources of CP violation and the SM sphalerons violate baryon+lepton
number. If HNLs are out of thermal equilibrium at any temperature higher
than that of sphaleron decoupling --- 130 GeV approximately for the 
SM\citep{DOnofrio:2014rug} --- then all of Sakharov's conditions for 
successful baryogenesis are satisfied. 
Traditionally it was believed that leptogenesis requires very large Majorana masses for the HNLs\citep{Davidson:2002qv}.\footnote{See\citep{Buchmuller:2005eh,Davidson:2008bu,Bodeker:2020ghk,DiBari:2021fhs} for reviews on high scale leptogenesis and\citep{Dev:2017trv,Drewes:2017zyw,Dev:2017wwc,Biondini:2017rpb,Chun:2017spz,Hagedorn:2017wjy} for reviews on individual aspects of both, high and low scale leptogenesis.} 
In such high scale scenarios the HNLs are inaccessible to direct searches, though the mechanism can in principle be falsified\citep{Deppisch:2013jxa} by combining collider data with searches for neutrinoless double $\beta$-decay ($0\nu\beta\beta$)\citep{Deppisch:2015yqa}, cf.~Sec.~\ref{sec:HighScaleLepto}.
However, it is now established that HNLs with masses below the TeV scale can generate the observed BAU during both their production\citep{Akhmedov:1998qx,Asaka:2005pn} and/or freeze-out and decay\citep{Pilaftsis:2003gt,Pilaftsis:2005rv, Dev:2014pfm}. These mechanisms can be embedded in theoretically well-motivated low scale seesaw scenarios\citep{Agrawal:2021dbo}  and open up the possibility to search for the HNLs at colliders\citep{Atre:2009rg,Deppisch:2015qwa,Cai:2017mow,Agrawal:2021dbo} and probe the origin of matter\citep{Chun:2017spz}. 

\subsection{Low scale leptogenesis mechanisms}

The mechanisms driving low scale leptogenesis can be classified by the dominant cause of the deviation from equilibrium, namely the \emph{freeze-out} and \emph{freeze-in} mechanisms. 
In the former case, HNLs attain first thermal equilibrium early
on, to then fall out of it --- freeze out --- while sphalerons are still active.
They will then decay out of equilibrium, with the CP violation
of the Yukawas compounded by the mass degeneracy to give rise to
\emph{resonant leptogenesis}\citep{Liu:1993tg,Flanz:1994yx,Flanz:1996fb,Covi:1996wh,Covi:1996fm,Pilaftsis:1997jf,Buchmuller:1997yu,Pilaftsis:2003gt}. 
In the latter case, Yukawas are small enough that the HNLs cannot manage
to equilibrate before sphaleron freeze-out.
It was long assumed that the freeze-out mechanism requires Majorana masses above the electroweak scale and the freeze-in mechanism only works with Majorana masses below the electroweak scale.
However, it has been recently pointed out\citep{Klaric:2020phc,Klaric:2021cpi}
that both mechanism in general coexist, and the assumption that one of them dominates only holds for $M \gg 1$ TeV\citep{Klaric:2020phc} or $M< 2$ GeV\citep{Drewes:2021nqr}. 

Low scale leptogenesis is technically more complicated than most high scale scenarios, due to thermal corrections to quasiparticle properties and reaction rates, the interplay between coherent oscillations and decoherent scatterings and a full flavour dependence. 
From the theoretical standpoint, computations of baryon asymmetry
production rely on techniques ranging from simple semi-classical Boltzmann
equations to more sophisticated approaches, such as the closed
time path (CTP) formalism or the operatorial approach based 
on scale separations, have been developed. We refer to these recent reviews\citep{Garbrecht:2018mrp,Biondini:2017rpb} for an
in-depth analysis of computational techniques. 
These have been used for both, precision calculations\citep{Ghiglieri:2018wbs,Ghiglieri:2020ulj}
as well as explorations of the viable parameters space \citep{Canetti:2010aw,Canetti:2012vf,Canetti:2012kh,Dev:2014oar, Hernandez:2015wna,Shuve:2014zua,Drewes:2016gmt,Hernandez:2016kel,Drewes:2016jae,Eijima:2018qke,Boiarska:2019jcw,Drewes:2021nqr,Klaric:2021cpi}.
The allowed parameter space overlaps with the sensitivity of many direct search experiments, see figure \ref{HNLsummary}.

\subsection{Testability and complementarity}

\begin{figure}[t!]
\begin{center}
\includegraphics[width=0.415\textwidth]{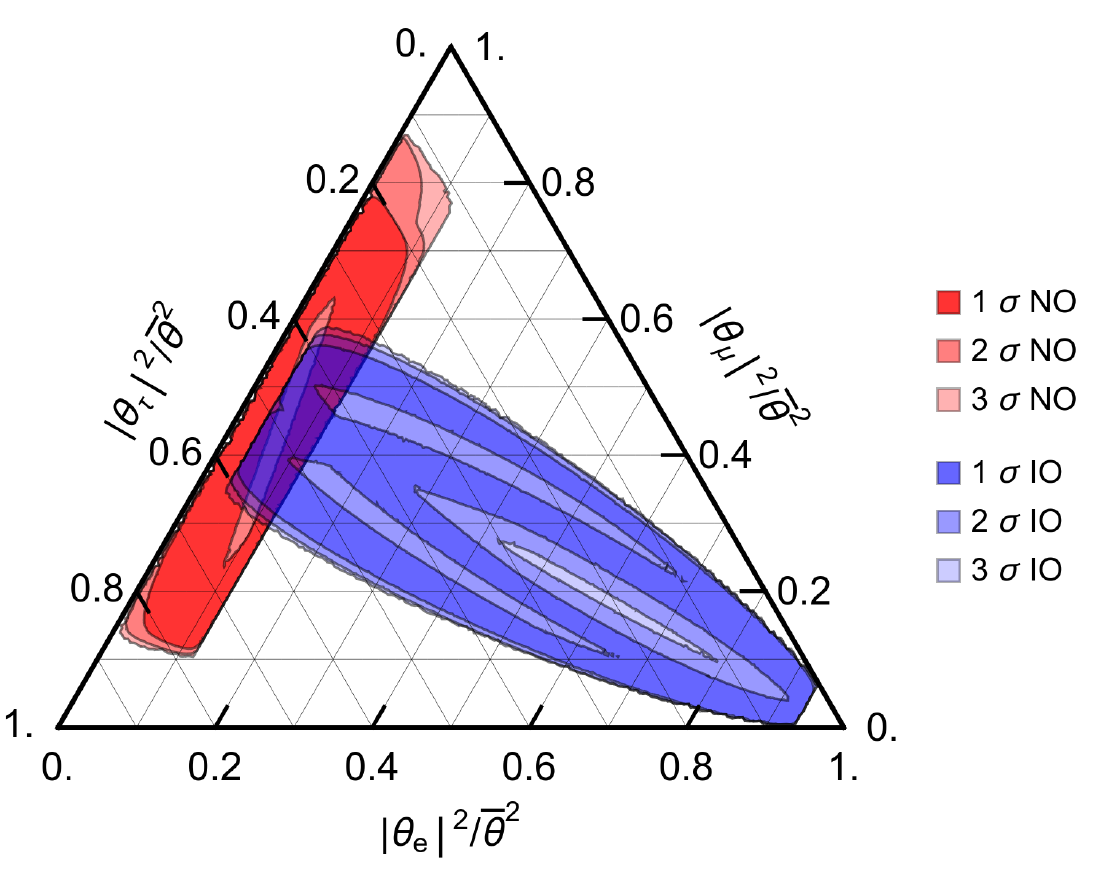}\includegraphics[width=0.45\textwidth]{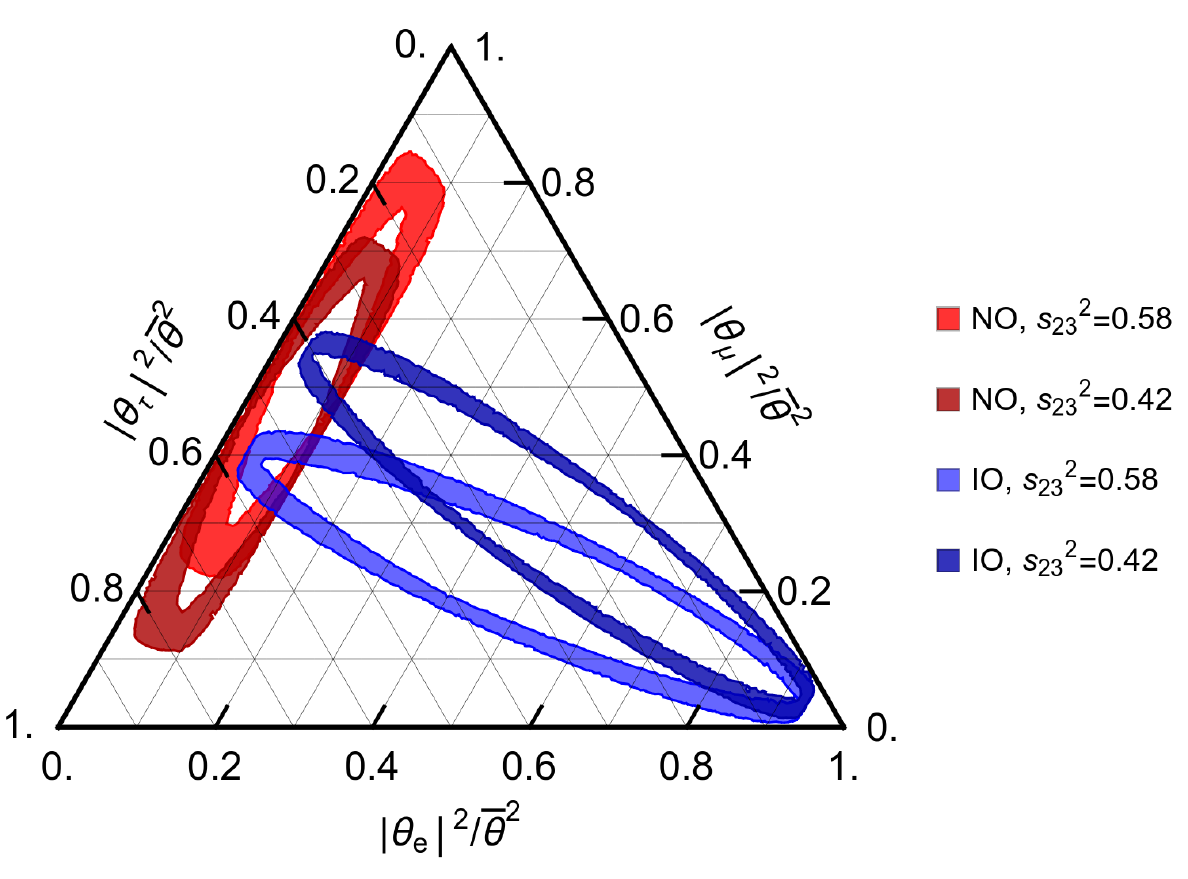}
\caption{\label{fig:triangle}
Allowed range for the relative magnitude of the HNL mixings with indicifual SM flavours in the minimal type I seesaw model\citep{Verhaaren:2022ify}.
 \textit{Left panel}: The range of normalized flavour mixings consistent with the current $\nu$ oscillation data 
 (see e.g.\citep{Drewes:2016jae,Caputo:2017pit,Drewes:2018gkc}). The different contours correspond to the allowed $\Delta \chi^2$ range from\citep{Esteban:2020cvm} for the case of normal (red) and inverted (blue) light $\nu$ mass ordering.
\textit{Right panel}: The projected $90\%$ CL for the mixing ratios after 14 years of $\nu$ oscillation measurements at DUNE\citep{DUNE:2020jqi}.
For this projection we assume maximal CP violation $\delta=-\pi/2$ and two benchmark values of the PMNS angle $\theta_{23}$, used in the DUNE TDR\citep{DUNE:2020ypp}, indicated in the legend.
}
\end{center}
\end{figure}

\begin{figure*}
\centering
\includegraphics[width=0.49\textwidth]{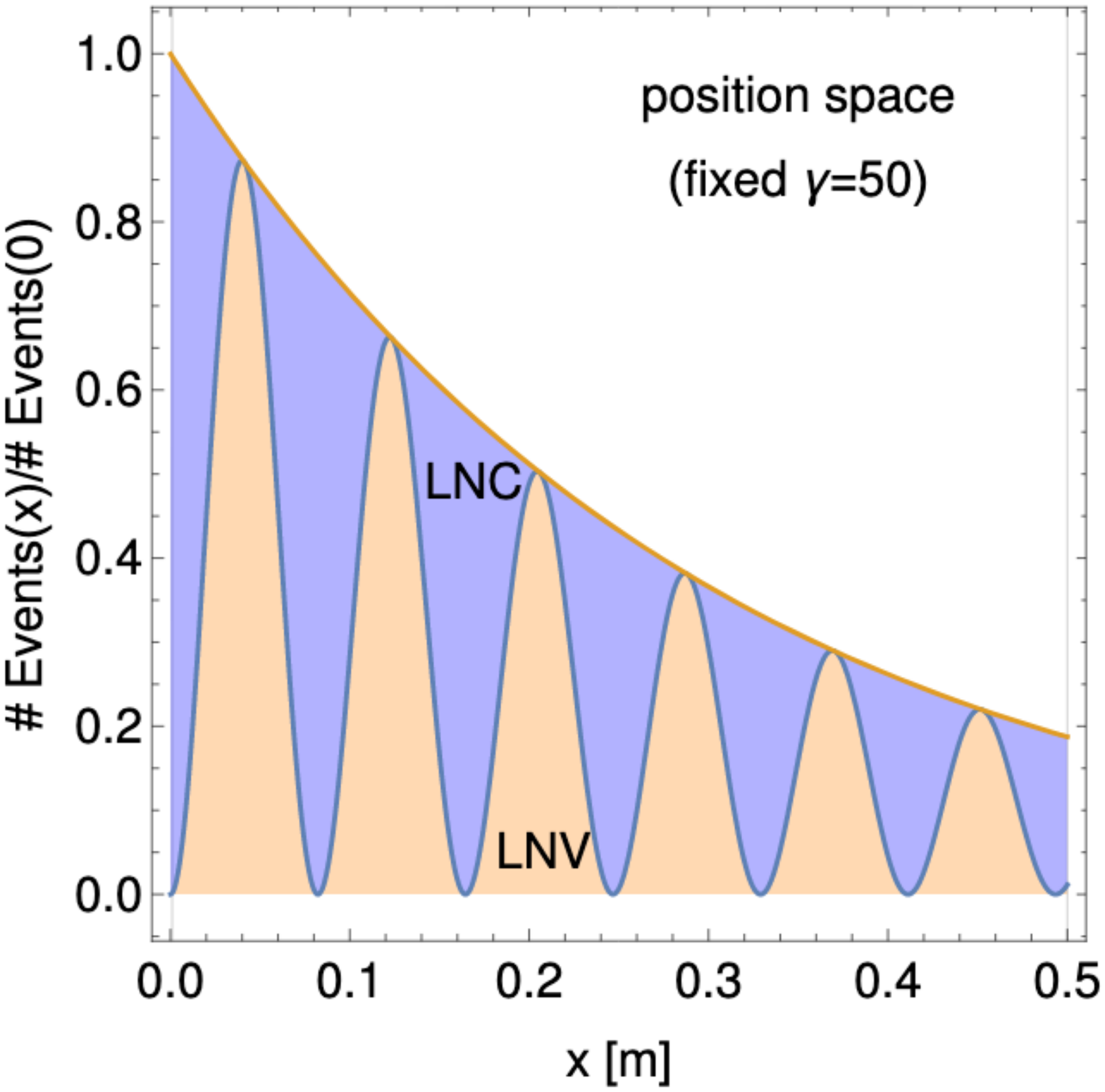}
\includegraphics[width=0.49\textwidth]{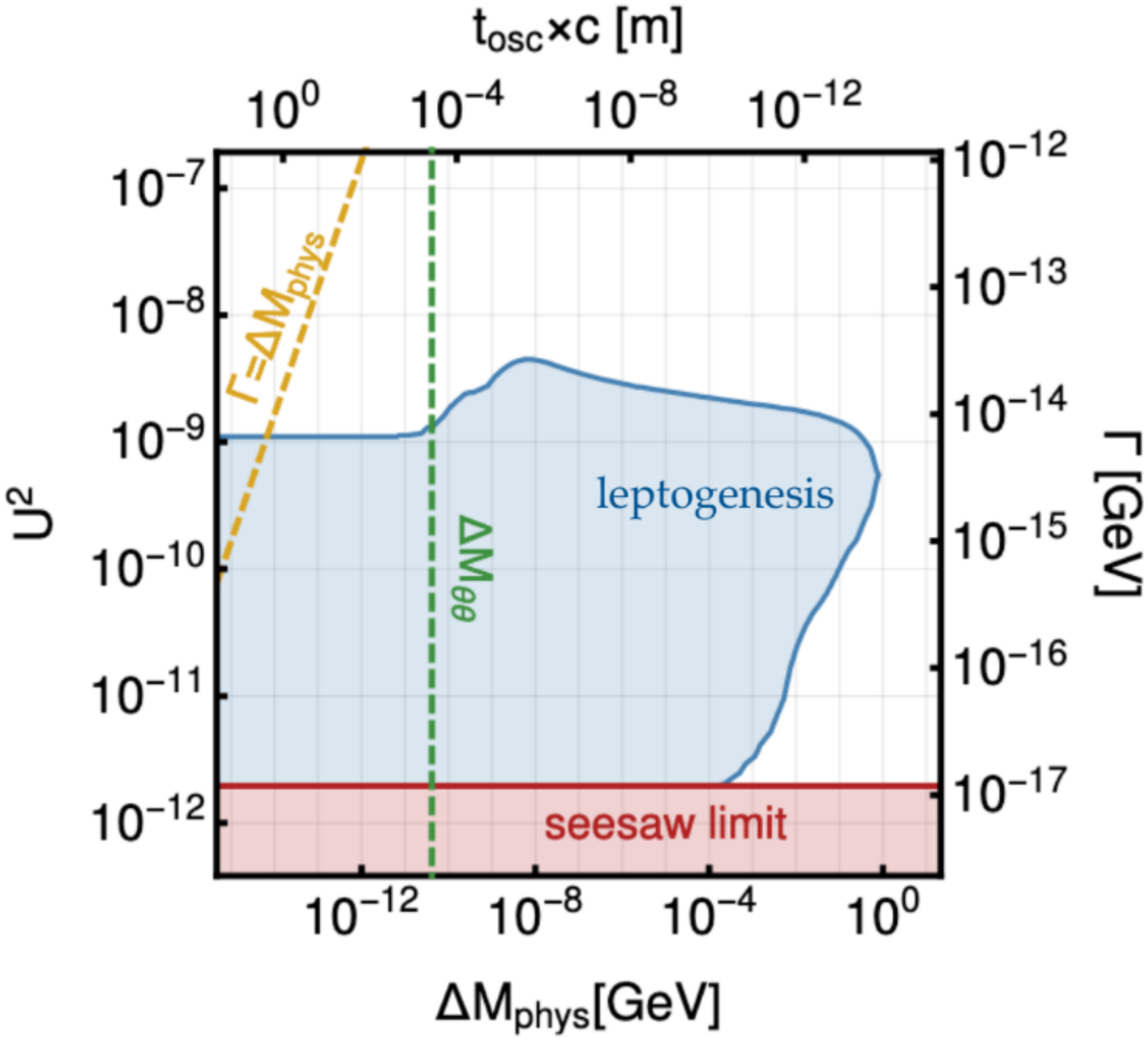}
\caption{\label{leptogenesisn2}
Left: Idealised picture of the ratio of $L$-violating (LNV) and $L$-conserving (LNC) heavy $\nu$ decays as a function of position. The ratio oscillates due to coherent $N_i$ flavour oscillations\citep{Antusch:2017ebe}.
Right: Range of physical heave $\nu$ mass splitting $\Delta M_{\rm phys}$ and mixing $U^2=\sum_iU_i^2$ for which leptogenesis is feasible for $\bar{M}=30$ GeV (blue region), compared to the contribution $\Delta M_{\theta\theta}$ to $\Delta M_{\rm phys}$ from the Higgs mechanism\citep{Antusch:2017pkq}.
}
\end{figure*}

\begin{figure}[t!]
 \begin{center}
 \includegraphics[width=0.45\columnwidth]{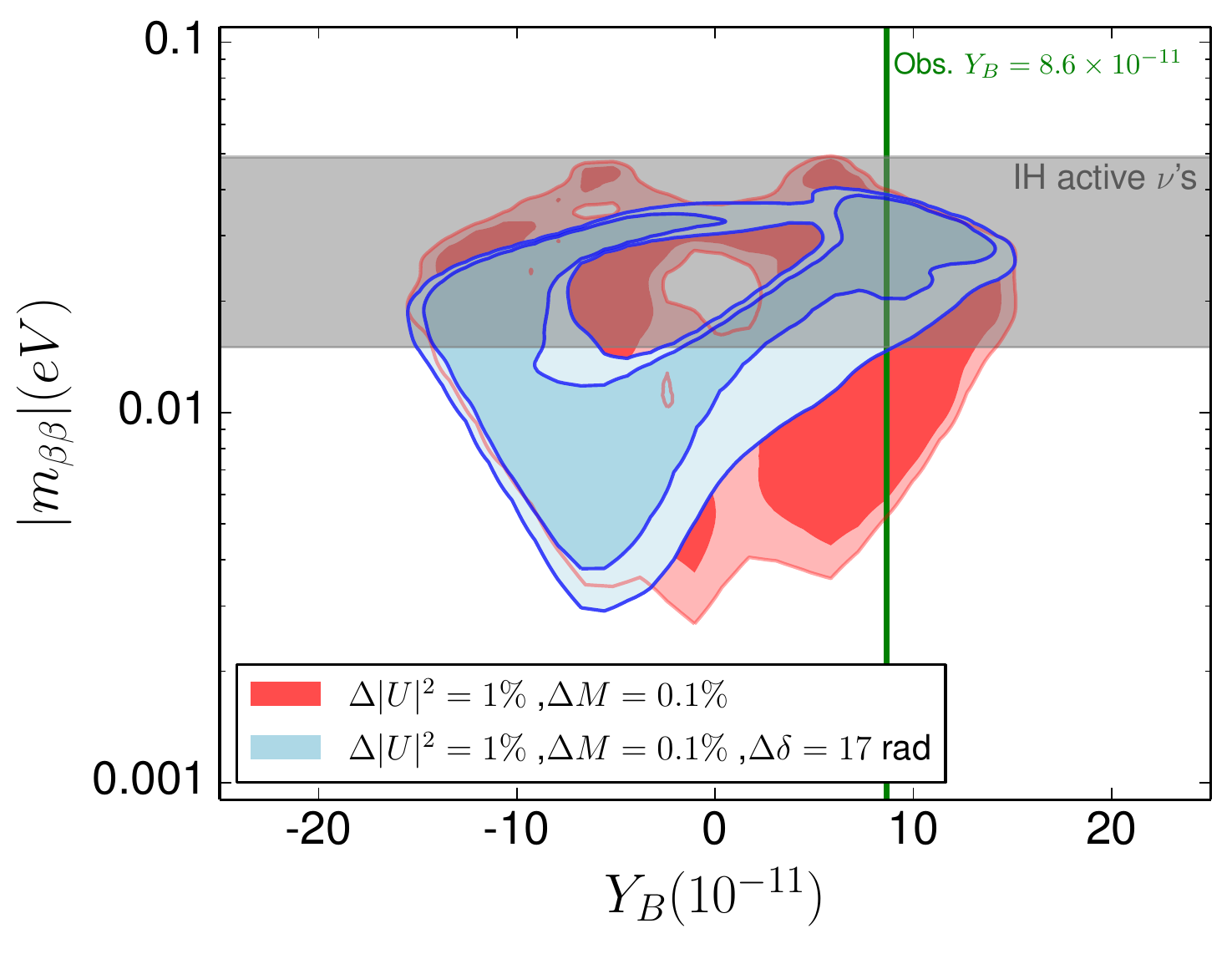} 
  \includegraphics[width=0.35\columnwidth]{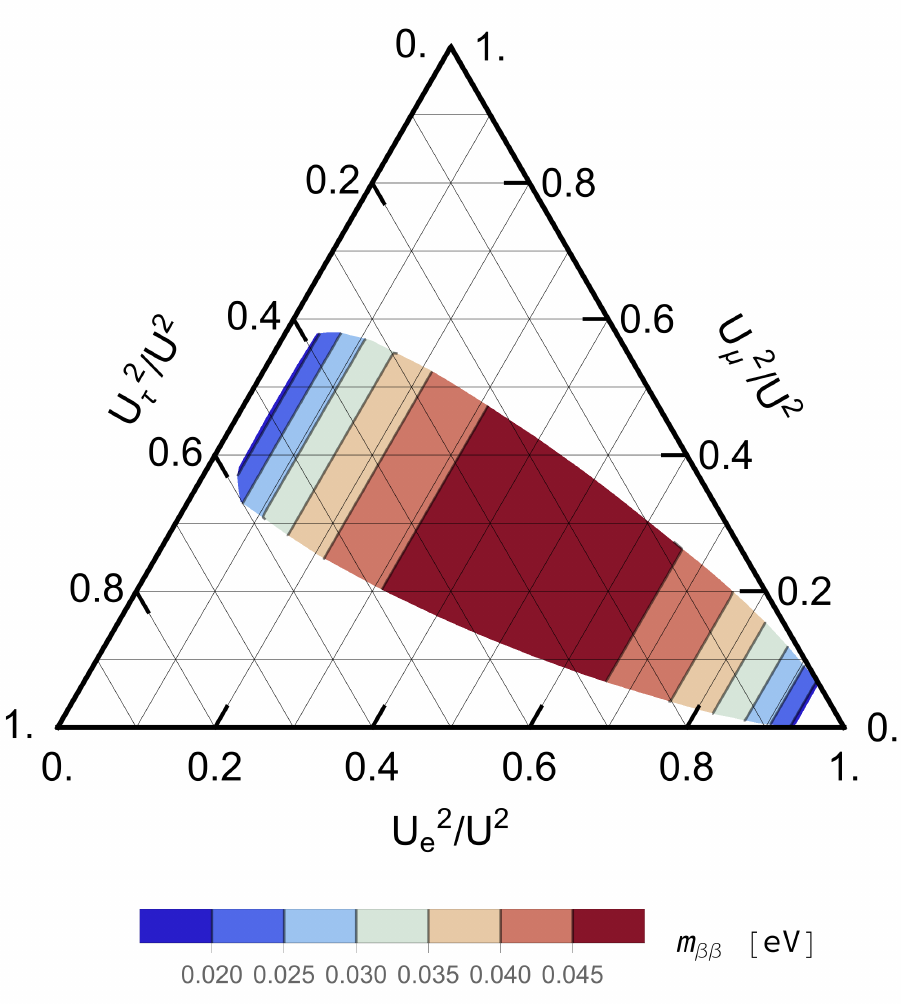} 
\caption{\label{fig:masterplot} \emph{Left panel:} 
Posterior probabilities in the $|m_{\beta\beta}|$ vs baryon asymmetry $Y_B$ plane 
in the type I seesaw model with two HNL flavours.
after a measurement at SHiP (red), assuming
$0.1\%, 1\%$ uncertainty on the measurement of the HNL masses and mixing respectively, and the an measurement of $\delta$ by T2HK or 
DUNE (blue). See\citep{Hernandez:2016kel} for further details.
\color{black}
\emph{Right panel:}
Expected rate for $0\nu\beta\beta$ in the minimal type I seesaw with two right-handed $\nu$s (or the $\nu$MSM) in the allowed region from Fig.~\ref{fig:triangle} for inverted light $\nu$ mass ordering. A deviation from this prediction would be sensitive to the heavy $\nu$ mass splitting and the  phases in the sterile sector\citep{Asaka:2016zib}, an important ingredient to test leptogenesis\citep{Drewes:2016lqo,Hernandez:2016kel,Asaka:2016zib}. 
}
\end{center}
\end{figure}

\begin{figure*}
 \begin{center}
\includegraphics[width=0.5\textwidth]{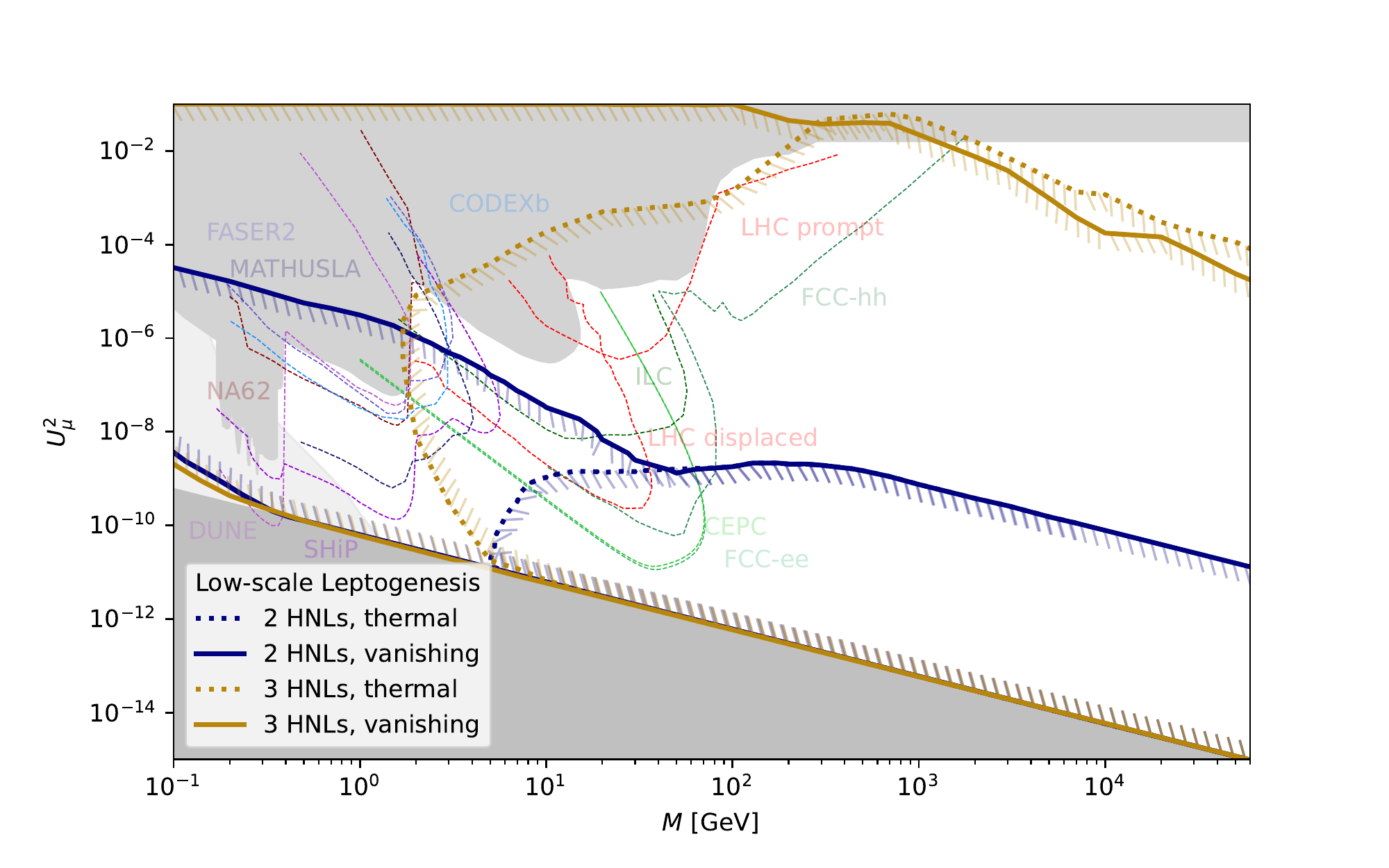}
\includegraphics[width=0.45\textwidth]{LNVvsLNC.pdf}
\caption
{
\emph{Left panel:}
Dark gray: Lower bound on the total HNL mixing from the requirement to explain the light $\nu$ oscillation data\citep{Esteban:2020cvm}.
Medium gray: Constraints on the active-sterile mixing $U_{\mu i}^2=\sum_\alpha|\theta_{\alpha i}|^2$ of HNLs from past experiments\citep{CHARM:1985nku,Abela:1981nf,Yamazaki:1984sj,E949:2014gsn,Bernardi:1987ek,NuTeV:1999kej,Vaitaitis:2000vc,CMS:2018iaf,DELPHI:1996qcc,ATLAS:2019kpx,CMS:2022fut}, obtained under the assumption that the HNLs exclusively mix with the second SM generation. 
Light gray: Lower bound on $U_\mu^2$ from BBN\citep{Sabti:2020yrt,Boyarsky:2020dzc}. Hashed orange and violet lines: Regions in which the observed baryon asymmetry of the universe can be explained with two\citep{Klaric:2020phc,Klaric:2021cpi} or three\citep{Drewes:2021nqr} HNL flavours and different initial conditions, as explained in the legend.
Other colourful lines: Estimated sensitivities of the LHC main detectors (taken from\citep{Izaguirre:2015pga,Drewes:2019fou,Pascoli:2018heg}) and NA62\citep{Drewes:2018gkc} as well as the sensitivities of selected planned or proposed experiments (DUNE\citep{Ballett:2019bgd},FASER2\citep{FASER:2018eoc},		 SHiP\citep{SHiP:2018xqw,Gorbunov:2020rjx}, MATHUSLA\citep{Curtin:2018mvb}, Codex-b\citep{Aielli:2019ivi}) as well as FCC-ee/CEPC\citep{Verhaaren:2022ify}, ILC and FCC-hh\citep{Antusch:2016ejd}.
Plot from \cite{Abdullahi:2022jlv}.
\emph{Right panel:} Regions of mass- and mixing where $L$-violating processes are expected to be observable in the minimal type I seesaw model\citep{Drewes:2019byd}, quantified by the ratio $R_{ll}$ of $L$-violating to $L$-conserving HNL decays\citep{Anamiati:2016uxp}.
}\label{HNLsummary}
\end{center}
\end{figure*}

\color{black}
If HNLs are discovered in any direct search, a pressing question will be whether these particles are indeed responsible for the origin of $\nu$ masses and the BAU. For this purpose one can take advantage that the parameter space is constrained by a wide range of indirect observables\citep{Drewes:2013gca,Antusch:2014woa,Gorbunov:2014ypa,Drewes:2015iva, deGouvea:2015euy,Fernandez-Martinez:2016lgt,Drewes:2016jae,Chrzaszcz:2019inj, Bolton:2019pcu}. Some points include:
\begin{itemize}
\item The requirement to reproduce the observed pattern of light $\nu$ mixing imposes testable constraints on the relative size of the HNL couplings to individual SM flavors\citep{Hernandez:2016kel,Drewes:2016jae,Drewes:2018gkc,Chrzaszcz:2019inj}. These will improve in the future with DUNE, see  Fig.~\ref{fig:triangle}.
\item A confirmation of the Majorana nature of the HNLs can be achieved by observing lepton number violation (LNV) in their decays. The smallness of the light $\nu$ masses enforces an generalised $B-L$ symmetry on the viable parameter space of low scale seesaw scenarios\citep{Shaposhnikov:2006nn,Kersten:2007vk}, implying that HNLs with collider-accessible couplings with the SM must be organised in pseudo-Dirac pairs\citep{Moffat:2017feq}. The resulting (approximate) $B-L$ conservation in principle parametrically suppresses all LNV processes at colliders\citep{Kersten:2007vk,Ibarra:2010xw}, but can be overcome if the HNL lifetime exceeds their oscillation length\citep{Anamiati:2016uxp} through oscillations between the two HNLs\citep{Boyanovsky:2014una,Antusch:2020pnn} with different lepton number. 
Even if the final state cannot be fully reconstructed, the angular distribution of decay products can be\citep{Kayser:1982br,Balantekin:2018ukw,Tastet:2019nqj,Blondel:2021mss,deGouvea:2021rpa}; similarly, decays into different flavors\citep{Dib:2016wge} can be sensitive to the LNV.
These effects have been studied for the LHC\citep{Dib:2017iva,Das:2017hmg}, ILC\citep{Hernandez:2018cgc}, NA62\citep{Abada:2019bac} and SHiP\citep{Tastet:2019nqj}. Generally one can expect LNV to observable in searches for long-lived HNLs, while it is suppressed in prompt decays\citep{Drewes:2019byd}; see Fig.~\ref{HNLsummary}.
\item Measuring the HNL flavour mixing pattern can permit indirect constraints on the Majorana phase in the $\nu$ mixing matrix\citep{Drewes:2016jae,Caputo:2016ojx}, while oscillations can give access to other CPV phases in the HNL sector\citep{Cvetic:2015naa,Cvetic:2015ura}.
The phases in the sterile sector can also be accessed within the viable leptogenesis parameter space by combining measurements of the HNL couplings to individual SM flavours with light $\nu$ oscillation data and $0\nu\beta\beta$\citep{Drewes:2016lqo,Hernandez:2016kel,Asaka:2016zib}\footnote{See also\citep{Bolton:2019pcu} for a discussion of the complementarity between $0\nu\beta\beta$ decays and colliders in the type I seesaw.} (if the sterile $\nu$ has a Majorana mass); see Fig.~\ref{fig:masterplot}. 
\item Finally, if HNL oscillations within the detector can be resolved, this can be used to indirectly constrain the HNL mass splitting\citep{Antusch:2017ebe,Antusch:2017pkq,Cvetic:2018elt,Cvetic:2019rms}, a parameter that strongly affects the BAU due to the resonant enhancement\citep{Dev:2017wwc,Garbrecht:2018mrp}, cf.~Fig.~\ref{leptogenesisn2}.
\end{itemize}
In addition, the HNLs can also be searched for using the meson decays, beta decays, electroweak precision data, and various cosmological and astrophysical observables; for a summary, see e.g.\citep{Bolton:2019pcu}. 

In the minimal realisation of the low scale seesaw, which phenomenologically also descibes the $\nu$MSM\citep{Asaka:2005an,Asaka:2005pn}, the parameter space is sufficiently constrained to fully test the model (in the sense of constraining all parameters in the Lagrangian) by combining all of these observables\citep{Hernandez:2016kel,Drewes:2016jae}.
Studies in a broader class of models will help to assess the potential of different experiments to discover the HNLs and, if they are found, probe the hypothesis that they are responsible for the origins of $\nu$ masses and the BAU.




\section{Probing leptogenesis with TeV-scale lepton-number violation}\label{sec:HighScaleLepto} 
The testability and complementarity of different experimental frontiers make low-scale scale leptogenesis an interesting scenario. However, leptogenesis could be similarly realised via a high-scale leptogenesis mechanism, for instance the out-of-equilibrium decay of right-handed neutrinos (RHNs) featuring CP-violating interactions. However, as the expected RHN masses\citep{Davidson:2002qv} are typically beyond the reach of current and future colliders (with the exception of e.g. resonant leptogenesis\citep{Pilaftsis:2003gt}), these scenarios are usually difficult to probe. In order to ultimately experimentally pinpoint the mechanism behind the baryon asymmetry, alternative approaches have to be identified in order to also probe models at high scales. One powerful possibility is to instead "falsify" high-scale leptogenesis models via the observation of lepton-number violation around the TeV-scale.

For instance, the observation of a same-sign dilepton signature without missing energy at the LHC, would imply such a strong washout that an asymmetry generated at a higher scale would be directly washed out, ruling out the possibility to have generated the baryon asymmetry via high-scale leptogenesis models\citep{Deppisch:2013jxa}. For leptogenesis scenarios around the TeV-scale, such as resonant leptogenesis, an observation of such a signal would imply a lower limit on the CP asymmetry.

Another lepton-number violating observable is $0\nu\beta\beta$ decay. If a new physics mechanism of $0\nu\beta\beta$ decay other than the standard light $\nu$-exchange via the Weinberg operator is observed, typical scenarios of high-scale leptogenesis will be excluded unless the baryon asymmetry is protected via some new mechanism\citep{Deppisch:2015yqa,Deppisch:2017ecm}. Hence, once $0\nu\beta\beta$ decay is observed, the identification of its underlying mechanism will be of great importance. In order to confirm washout in all flavour sectors, an observation in all flavours or an additional measurement of lepton-flavour violation is required.

The interplay between collider probes and $0\nu\beta\beta$ decay is particularly important. Depending on the hierarchy of the new UV physics involved, one or the other probe can have the larger experimental reach\citep{Harz:2021psp}. While $0\nu\beta\beta$ decay is limited to the first generation only, collider searches allow also for same-sign signatures in second or third generation leptons. A detailed analysis\citep{Harz:2021psp} of a simplified model based on a dim-9 effective LNV interaction confirmed that an observation of lepton-number violation at current or future colliders or at $0\nu\beta\beta$ decay experiments would render single-flavour standard high-scale leptogenesis invalid. Again, for a conclusive exclusion equilibration in all flavours has to be confirmed.

It is important to emphasize that these far-reaching consequences are not limited to high-scale leptogenesis scenarios, but similarly to other high-scale $\Delta(B-L) \neq 0$ baryogenesis scenarios due to the equilibration via the $(B+L)$-violating sphaleron processes. 
However, certain models that generate an asymmetry in a secluded dark sector or hide an asymmetry in a specific flavour might be exempt from such conclusions\citep{AristizabalSierra:2013lyx,Frandsen:2018jfi}.

In summary, experimental searches for lepton-number-violating signatures are highly relevant for probing high-scale leptogenesis and baryogenesis scenarios, even though the physics that generates the asymmetry might not be directly accessible. Hence, the observation of lepton-number violation at $0\nu\beta\beta$ decay experiments via a nonstandard mechanism or at current and future colliders, might point us to low-scale baryogenesis. Therefore, searches for lepton-number violation should be pursued with high priority at all frontiers.

%% file: main.bbl
\begin{thebibliography}{100}

\bibitem{Sakharov:1967dj}
A.~Sakharov,
\newblock Sov. Phys. Usp. {\bf 34}, 392 (1991).

\bibitem{tHooft:1976snw}
G.~'t~Hooft,
\newblock Phys. Rev. D {\bf 14}, 3432 (1976),
\newblock [Erratum: Phys.Rev.D 18, 2199 (1978)].

\bibitem{Dolgov:1991fr}
A.~Dolgov,
\newblock Phys. Rept. {\bf 222}, 309 (1992).

\bibitem{Babu:2006xc}
K.~S. Babu, R.~N. Mohapatra, and S.~Nasri,
\newblock Phys. Rev. Lett. {\bf 97}, 131301 (2006), hep-ph/0606144.

\bibitem{NelsonINT2017}
A.~Nelson,
\newblock {CP Violation, Baryon violation, RPV in SUSY, Mesino Oscillations,
  and Baryogenesis}, 2017,
\newblock Slide 4:
  int.washington.edu/talks/WorkShops/int\_17\_69W/People/Nelson\_A/Nelson.pdf.

\bibitem{Pati:1974yy}
J.~C. Pati and A.~Salam,
\newblock Phys. Rev. D {\bf 10}, 275 (1974),
\newblock [Erratum: Phys.Rev.D 11, 703--703 (1975)].

\bibitem{Babu:2008rq}
K.~Babu, P.~S.~B. Dev, and R.~Mohapatra,
\newblock Phys. Rev. D {\bf 79}, 015017 (2009), 0811.3411.

\bibitem{Babu:2013yca}
K.~S. Babu, P.~S.~B. Dev, E.~C. F.~S. Fortes, and R.~N. Mohapatra,
\newblock Phys. Rev. {\bf D87}, 115019 (2013), 1303.6918.

\bibitem{BaldoCeolin:1994jz}
M.~Baldo-Ceolin {\em et~al.},
\newblock Z. Phys. C {\bf 63}, 409 (1994).

\bibitem{Super-Kamiokande:2020bov}
Super-Kamiokande, K.~Abe {\em et~al.},
\newblock Phys. Rev. D {\bf 103}, 012008 (2021), 2012.02607.

\bibitem{Rinaldi:2018osy}
E.~Rinaldi {\em et~al.},
\newblock Phys. Rev. Lett. {\bf 122}, 162001 (2019), 1809.00246.

\bibitem{Rinaldi:2019thf}
E.~Rinaldi {\em et~al.},
\newblock Phys. Rev. D {\bf 99}, 074510 (2019), 1901.07519.

\bibitem{Rao:1982gt}
S.~Rao and R.~Shrock,
\newblock Phys. Lett. B {\bf 116}, 238 (1982).

\bibitem{BhupalPresentation}
P.~S.~B. Dev,
\newblock {Update on the post-sphaleron baryogenesis model prediction for
  neutron-antineutron oscillation time}, 2020,
\newblock
  {\href{https://indico.fnal.gov/event/44472/contributions/192053/}{Link
  here}}.

\bibitem{Proceedings:2020nzz}
{\em {$|\Delta \mathcal{B}| =2$: A State of the Field, and Looking Forward--A
  brief status report of theoretical and experimental physics opportunities}},
  2020, 2010.02299.

\bibitem{Mohapatra:2007af}
R.~N. Mohapatra, N.~Okada, and H.-B. Yu,
\newblock Phys. Rev. D {\bf 77}, 011701 (2008), 0709.1486.

\bibitem{Chen:2008hh}
C.-R. Chen, W.~Klemm, V.~Rentala, and K.~Wang,
\newblock Phys. Rev. D {\bf 79}, 054002 (2009), 0811.2105.

\bibitem{Berger:2010fy}
E.~L. Berger, Q.-H. Cao, C.-R. Chen, G.~Shaughnessy, and H.~Zhang,
\newblock Phys. Rev. Lett. {\bf 105}, 181802 (2010), 1005.2622.

\bibitem{Baldes:2011mh}
I.~Baldes, N.~F. Bell, and R.~R. Volkas,
\newblock Phys. Rev. D {\bf 84}, 115019 (2011), 1110.4450.

\bibitem{Chivukula:2015zma}
R.~S. Chivukula, P.~Ittisamai, K.~Mohan, and E.~H. Simmons,
\newblock Phys. Rev. D {\bf 92}, 075020 (2015), 1507.06676.

\bibitem{CMS:2019gwf}
CMS, A.~M. Sirunyan {\em et~al.},
\newblock JHEP {\bf 05}, 033 (2020), 1911.03947.

\bibitem{Pascual-Dias:2020hxo}
B.~Pascual-Dias, P.~Saha, and D.~London,
\newblock JHEP {\bf 07}, 144 (2020), 2006.13385.

\bibitem{Bell:2018mgg}
N.~F. Bell, T.~Corbett, M.~Nee, and M.~J. Ramsey-Musolf,
\newblock Phys. Rev. D {\bf 99}, 015034 (2019), 1808.10597.

\bibitem{Dev:2015uca}
P.~S.~B. Dev and R.~N. Mohapatra,
\newblock Phys. Rev. D {\bf 92}, 016007 (2015), 1504.07196.

\bibitem{Allahverdi:2017edd}
R.~Allahverdi, P.~S.~B. Dev, and B.~Dutta,
\newblock Phys. Lett. B {\bf 779}, 262 (2018), 1712.02713.

\bibitem{Kolb:1979qa}
E.~W. Kolb and S.~Wolfram,
\newblock Nucl. Phys. B {\bf 172}, 224 (1980),
\newblock [Erratum: Nucl.Phys.B 195, 542 (1982)].

\bibitem{Allahverdi:2013mza}
R.~Allahverdi and B.~Dutta,
\newblock Phys. Rev. D {\bf 88}, 023525 (2013), 1304.0711.

\bibitem{Cui:2012jh}
Y.~Cui and R.~Sundrum,
\newblock Phys. Rev. D {\bf 87}, 116013 (2013), 1212.2973.

\bibitem{Cui:2013bta}
Y.~Cui,
\newblock JHEP {\bf 12}, 067 (2013), 1309.2952.

\bibitem{Arcadi:2015ffa}
G.~Arcadi, L.~Covi, and M.~Nardecchia,
\newblock Phys. Rev. D {\bf 92}, 115006 (2015), 1507.05584.

\bibitem{Pierce:2019ozl}
A.~Pierce and B.~Shakya,
\newblock JHEP {\bf 06}, 096 (2019), 1901.05493.

\bibitem{Grojean:2018fus}
C.~Grojean, B.~Shakya, J.~D. Wells, and Z.~Zhang,
\newblock Phys. Rev. Lett. {\bf 121}, 171801 (2018), 1806.00011.

\bibitem{Fridell:2021gag}
K.~Fridell, J.~Harz, and C.~Hati,
\newblock JHEP {\bf 11}, 185 (2021), 2105.06487.

\bibitem{Zwirner:1984is}
F.~Zwirner,
\newblock Phys. Lett. B {\bf 132}, 103 (1983).

\bibitem{Barbieri:1985ty}
R.~Barbieri and A.~Masiero,
\newblock Nucl. Phys. B {\bf 267}, 679 (1986).

\bibitem{Mohapatra:1986bd}
R.~Mohapatra and J.~Valle,
\newblock Phys. Rev. D {\bf 34}, 1642 (1986).

\bibitem{Lazarides:1986jt}
G.~Lazarides, C.~Panagiotakopoulos, and Q.~Shafi,
\newblock Phys. Lett. B {\bf 175}, 309 (1986).

\bibitem{Goity:1994dq}
J.~Goity and M.~Sher,
\newblock Phys. Lett. B {\bf 346}, 69 (1995), hep-ph/9412208,
\newblock [Erratum: Phys.Lett.B 385, 500 (1996)].

\bibitem{Babu:2001qr}
K.~Babu and R.~Mohapatra,
\newblock Phys. Lett. B {\bf 518}, 269 (2001), hep-ph/0108089.

\bibitem{Babu:2006wz}
K.~Babu, R.~Mohapatra, and S.~Nasri,
\newblock Phys. Rev. Lett. {\bf 98}, 161301 (2007), hep-ph/0612357.

\bibitem{Allahverdi:2010im}
R.~Allahverdi, B.~Dutta, and K.~Sinha,
\newblock Phys. Rev. D {\bf 82}, 035004 (2010), 1005.2804.

\bibitem{Gu:2011ff}
P.-H. Gu and U.~Sarkar,
\newblock Phys. Lett. B {\bf 705}, 170 (2011), 1107.0173.

\bibitem{Gu:2011fp}
P.-H. Gu and U.~Sarkar,
\newblock (2011), 1110.2926.

\bibitem{Dhuria:2015swa}
M.~Dhuria, C.~Hati, and U.~Sarkar,
\newblock Phys. Rev. D {\bf 93}, 015001 (2016), 1507.08297.

\bibitem{Ghalsasi:2015mxa}
A.~Ghalsasi, D.~McKeen, and A.~E. Nelson,
\newblock Phys. Rev. D {\bf 92}, 076014 (2015), 1508.05392.

\bibitem{Gu:2016ghu}
P.-H. Gu, E.~Ma, and U.~Sarkar,
\newblock Phys. Rev. D {\bf 94}, 111701 (2016), 1608.02118.

\bibitem{Calibbi:2016ukt}
L.~Calibbi, G.~Ferretti, D.~Milstead, C.~Petersson, and R.~Pöttgen,
\newblock JHEP {\bf 05}, 144 (2016), 1602.04821,
\newblock [Erratum: JHEP 10, 195 (2017)].

\bibitem{Gu:2017cgp}
P.-H. Gu and U.~Sarkar,
\newblock Phys. Rev. D {\bf 96}, 031703 (2017), 1705.02858.

\bibitem{Calibbi:2017rab}
L.~Calibbi, E.~J. Chun, and C.~S. Shin,
\newblock JHEP {\bf 10}, 177 (2017), 1708.06439.

\bibitem{Mohapatra:1980qe}
R.~N. Mohapatra and R.~Marshak,
\newblock Phys. Rev. Lett. {\bf 44}, 1316 (1980),
\newblock [Erratum: Phys.Rev.Lett. 44, 1643 (1980)].

\bibitem{Mohapatra:1982xz}
R.~N. Mohapatra and G.~Senjanovic,
\newblock Phys. Rev. D {\bf 27}, 254 (1983).

\bibitem{Chang:1984qr}
D.~Chang, R.~Mohapatra, J.~Gipson, R.~Marshak, and M.~Parida,
\newblock Phys. Rev. D {\bf 31}, 1718 (1985).

\bibitem{Babu:2012vc}
K.~Babu and R.~Mohapatra,
\newblock Phys. Lett. B {\bf 715}, 328 (2012), 1206.5701.

\bibitem{Arnold:2012sd}
J.~M. Arnold, B.~Fornal, and M.~B. Wise,
\newblock Phys. Rev. D {\bf 87}, 075004 (2013), 1212.4556.

\bibitem{Patra:2014goa}
S.~Patra and P.~Pritimita,
\newblock Eur. Phys. J. C {\bf 74}, 3078 (2014), 1405.6836.

\bibitem{Herrmann:2014fha}
E.~Herrmann,
\newblock (2014), 1408.4455.

\bibitem{Nussinov:2001rb}
S.~Nussinov and R.~Shrock,
\newblock Phys. Rev. Lett. {\bf 88}, 171601 (2002), hep-ph/0112337.

\bibitem{Girmohanta:2019fsx}
S.~Girmohanta and R.~Shrock,
\newblock Phys. Rev. D {\bf 101}, 015017 (2020), 1911.05102.

\bibitem{Phillips:2014fgb}
I.~Phillips, D.G. {\em et~al.},
\newblock Phys. Rept. {\bf 612}, 1 (2016), 1410.1100.

\bibitem{Addazi:2020nlz}
A.~Addazi {\em et~al.},
\newblock (2020), 2006.04907.

\bibitem{BITTER1985461}
T.~Bitter and D.~Dubbers,
\newblock Nuclear Instruments and Methods in Physics Research Section A:
  Accelerators, Spectrometers, Detectors and Associated Equipment {\bf 239},
  461  (1985).

\bibitem{Davis:2016uyk}
E.~D. Davis and A.~R. Young,
\newblock Phys. Rev. {\bf D95}, 036004 (2017), 1611.04205.

\bibitem{Klinkby:2014cma}
E.~Klinkby {\em et~al.},
\newblock (2014), 1401.6003.

\bibitem{Santoro:2020nke}
V.~Santoro {\em et~al.},
\newblock Journal of Neutron Research {\bf Pre-press}, 1 (2020), 2002.03883.

\bibitem{Soldner:2018ycf}
T.~Soldner {\em et~al.},
\newblock EPJ Web Conf. {\bf 219}, 10003 (2019), 1811.11692.

\bibitem{Berezhiani:2005hv}
Z.~Berezhiani and L.~Bento,
\newblock Phys. Rev. Lett. {\bf 96}, 081801 (2006), hep-ph/0507031.

\bibitem{Foot:2004pa}
R.~Foot,
\newblock Int. J. Mod. Phys. {\bf D13}, 2161 (2004), astro-ph/0407623.

\bibitem{Foot:2014mia}
R.~Foot,
\newblock Int. J. Mod. Phys. {\bf A29}, 1430013 (2014), 1401.3965.

\bibitem{Berezhiani:2020nzn}
Z.~Berezhiani,
\newblock (2020), 2002.05609.

\bibitem{Hewes:2017xtr}
J.~E.~T. Hewes,
\newblock {\em {Searches for Bound Neutron-Antineutron Oscillation in Liquid
  Argon Time Projection Chambers}},
\newblock PhD thesis, Manchester U., 2017.

\bibitem{Golubeva:1997fs}
E.~Golubeva and L.~Kondratyuk,
\newblock Nucl. Phys. B Proc. Suppl. {\bf 56}, 103 (1997).

\bibitem{Golubeva:2018mrz}
E.~S. Golubeva, J.~L. Barrow, and C.~G. Ladd,
\newblock Phys. Rev. {\bf D99}, 035002 (2019), 1804.10270.

\bibitem{Barrow:2019viz}
J.~L. Barrow, E.~S. Golubeva, E.~Paryev, and J.-M. Richard,
\newblock Phys. Rev. D {\bf 101}, 036008 (2020), 1906.02833.

\bibitem{mezei1976novel}
F.~Mezei,
\newblock Communications on Physics (London) {\bf 1}, 81 (1976).

\bibitem{Granada:2020ksc}
J.~Granada, J.~M. Damián, and C.~Helman,
\newblock EPJ Web Conf. {\bf 231}, 04002 (2020).

\bibitem{Jamalipour:2020lsp}
M.~Jamalipour, L.~Zanini, and G.~Gorini,
\newblock EPJ Web Conf. {\bf 231}, 04003 (2020).

\bibitem{Abele:2005xd}
H.~Abele {\em et~al.},
\newblock Nucl. Instrum. Meth. A {\bf 562}, 407 (2006), nucl-ex/0510072.

\bibitem{FrostPhD}
M.~J. Frost,
\newblock {\em {Searching for Baryon Number Violation at Cold Neutron
  Sources}},
\newblock PhD thesis, The University of Tennessee at Knoxville, 2020,
\newblock Available upon request.

\bibitem{Manton:1983nd}
N.~S. Manton,
\newblock Phys. Rev. D {\bf 28}, 2019 (1983).

\bibitem{Klinkhamer:1984di}
F.~R. Klinkhamer and N.~S. Manton,
\newblock Phys. Rev. D {\bf 30}, 2212 (1984).

\bibitem{Bochkarev:1987wf}
A.~I. Bochkarev and M.~E. Shaposhnikov,
\newblock Mod. Phys. Lett. A {\bf 2}, 417 (1987).

\bibitem{Kajantie:1995kf}
K.~Kajantie, M.~Laine, K.~Rummukainen, and M.~E. Shaposhnikov,
\newblock Nucl. Phys. B {\bf 466}, 189 (1996), hep-lat/9510020.

\bibitem{Gavela:1993ts}
M.~B. Gavela, P.~Hernandez, J.~Orloff, and O.~Pene,
\newblock Mod. Phys. Lett. A {\bf 9}, 795 (1994), hep-ph/9312215.

\bibitem{Huet:1994jb}
P.~Huet and E.~Sather,
\newblock Phys. Rev. D {\bf 51}, 379 (1995), hep-ph/9404302.

\bibitem{Gavela:1994dt}
M.~B. Gavela, P.~Hernandez, J.~Orloff, O.~Pene, and C.~Quimbay,
\newblock Nucl. Phys. B {\bf 430}, 382 (1994), hep-ph/9406289.

\bibitem{Cohen:1992zx}
A.~G. Cohen and A.~E. Nelson,
\newblock Phys. Lett. B {\bf 297}, 111 (1992), hep-ph/9209245.

\bibitem{Carena:1996wj}
M.~Carena, M.~Quiros, and C.~E.~M. Wagner,
\newblock Phys. Lett. B {\bf 380}, 81 (1996), hep-ph/9603420.

\bibitem{Riotto:1997vy}
A.~Riotto,
\newblock Nucl. Phys. B {\bf 518}, 339 (1998), hep-ph/9712221.

\bibitem{Cline:1997vk}
J.~M. Cline, M.~Joyce, and K.~Kainulainen,
\newblock Phys. Lett. B {\bf 417}, 79 (1998), hep-ph/9708393,
\newblock [Erratum: Phys.Lett.B 448, 321--321 (1999)].

\bibitem{Cline:2000kb}
J.~M. Cline and K.~Kainulainen,
\newblock Phys. Rev. Lett. {\bf 85}, 5519 (2000), hep-ph/0002272.

\bibitem{Cline:2000nw}
J.~M. Cline, M.~Joyce, and K.~Kainulainen,
\newblock JHEP {\bf 07}, 018 (2000), hep-ph/0006119.

\bibitem{Lee:2004we}
C.~Lee, V.~Cirigliano, and M.~J. Ramsey-Musolf,
\newblock Phys. Rev. D {\bf 71}, 075010 (2005), hep-ph/0412354.

\bibitem{Chung:2009qs}
D.~J.~H. Chung, B.~Garbrecht, M.~J. Ramsey-Musolf, and S.~Tulin,
\newblock JHEP {\bf 12}, 067 (2009), 0908.2187.

\bibitem{Cirigliano:2006wh}
V.~Cirigliano, M.~J. Ramsey-Musolf, S.~Tulin, and C.~Lee,
\newblock Phys. Rev. D {\bf 73}, 115009 (2006), hep-ph/0603058.

\bibitem{Liebler:2015ddv}
S.~Liebler, S.~Profumo, and T.~Stefaniak,
\newblock JHEP {\bf 04}, 143 (2016), 1512.09172.

\bibitem{Turok:1990zg}
N.~Turok and J.~Zadrozny,
\newblock Nucl. Phys. B {\bf 358}, 471 (1991).

\bibitem{Dine:1990fj}
M.~Dine, P.~Huet, R.~L. Singleton, Jr, and L.~Susskind,
\newblock Phys. Lett. B {\bf 257}, 351 (1991).

\bibitem{Fromme:2006cm}
L.~Fromme, S.~J. Huber, and M.~Seniuch,
\newblock JHEP {\bf 11}, 038 (2006), hep-ph/0605242.

\bibitem{Liu:2011jh}
T.~Liu, M.~J. Ramsey-Musolf, and J.~Shu,
\newblock Phys. Rev. Lett. {\bf 108}, 221301 (2012), 1109.4145.

\bibitem{Kainulainen:2019kyp}
K.~Kainulainen {\em et~al.},
\newblock JHEP {\bf 06}, 075 (2019), 1904.01329.

\bibitem{Bian:2014zka}
L.~Bian, T.~Liu, and J.~Shu,
\newblock Phys. Rev. Lett. {\bf 115}, 021801 (2015), 1411.6695.

\bibitem{Fuyuto:2019svr}
K.~Fuyuto, W.-S. Hou, and E.~Senaha,
\newblock Phys. Rev. D {\bf 101}, 011901 (2020), 1910.12404.

\bibitem{Egana-Ugrinovic:2017jib}
D.~Egana-Ugrinovic,
\newblock JHEP {\bf 12}, 064 (2017), 1707.02306.

\bibitem{DeVries:2018aul}
J.~De~Vries, M.~Postma, and J.~van~de Vis,
\newblock JHEP {\bf 04}, 024 (2019), 1811.11104.

\bibitem{Cline:2012hg}
J.~M. Cline and K.~Kainulainen,
\newblock JCAP {\bf 01}, 012 (2013), 1210.4196.

\bibitem{Vaskonen:2016yiu}
V.~Vaskonen,
\newblock Phys. Rev. D {\bf 95}, 123515 (2017), 1611.02073.

\bibitem{Cheung:2012pg}
K.~Cheung, T.-J. Hou, J.~S. Lee, and E.~Senaha,
\newblock Phys. Lett. B {\bf 710}, 188 (2012), 1201.3781.

\bibitem{Bian:2017wfv}
L.~Bian, H.-K. Guo, and J.~Shu,
\newblock Chin. Phys. C {\bf 42}, 093106 (2018), 1704.02488,
\newblock [Erratum: Chin.Phys.C 43, 129101 (2019)].

\bibitem{Akula:2017yfr}
S.~Akula, C.~Bal\'azs, L.~Dunn, and G.~White,
\newblock JHEP {\bf 11}, 051 (2017), 1706.09898.

\bibitem{Bruggisser:2018mrt}
S.~Bruggisser, B.~Von~Harling, O.~Matsedonskyi, and G.~Servant,
\newblock JHEP {\bf 12}, 099 (2018), 1804.07314.

\bibitem{Ellis:2019flb}
S.~A.~R. Ellis, S.~Ipek, and G.~White,
\newblock JHEP {\bf 08}, 002 (2019), 1905.11994.

\bibitem{Cline:2017qpe}
J.~M. Cline, K.~Kainulainen, and D.~Tucker-Smith,
\newblock Phys. Rev. D {\bf 95}, 115006 (2017), 1702.08909.

\bibitem{Carena:2018cjh}
M.~Carena, M.~Quir\'os, and Y.~Zhang,
\newblock Phys. Rev. Lett. {\bf 122}, 201802 (2019), 1811.09719.

\bibitem{Baldes:2018nel}
I.~Baldes and G.~Servant,
\newblock JHEP {\bf 10}, 053 (2018), 1807.08770.

\bibitem{Glioti:2018roy}
A.~Glioti, R.~Rattazzi, and L.~Vecchi,
\newblock JHEP {\bf 04}, 027 (2019), 1811.11740.

\bibitem{Inoue:2015pza}
S.~Inoue, G.~Ovanesyan, and M.~J. Ramsey-Musolf,
\newblock Phys. Rev. D {\bf 93}, 015013 (2016), 1508.05404.

\bibitem{Ramsey-Musolf:2017tgh}
M.~J. Ramsey-Musolf, P.~Winslow, and G.~White,
\newblock Phys. Rev. D {\bf 97}, 123509 (2018), 1708.07511.

\bibitem{Anderson:1991zb}
G.~W. Anderson and L.~J. Hall,
\newblock Phys. Rev. D {\bf 45}, 2685 (1992).

\bibitem{Espinosa:2011ax}
J.~R. Espinosa, T.~Konstandin, and F.~Riva,
\newblock Nucl. Phys. B {\bf 854}, 592 (2012), 1107.5441.

\bibitem{Cline:2013gha}
J.~M. Cline, K.~Kainulainen, P.~Scott, and C.~Weniger,
\newblock Phys. Rev. D {\bf 88}, 055025 (2013), 1306.4710,
\newblock [Erratum: Phys.Rev.D 92, 039906 (2015)].

\bibitem{Curtin:2014jma}
D.~Curtin, P.~Meade, and C.-T. Yu,
\newblock JHEP {\bf 11}, 127 (2014), 1409.0005.

\bibitem{Huang:2016cjm}
P.~Huang, A.~J. Long, and L.-T. Wang,
\newblock Phys. Rev. D {\bf 94}, 075008 (2016), 1608.06619.

\bibitem{Cepeda:2019klc}
M.~Cepeda {\em et~al.},
\newblock CERN Yellow Rep. Monogr. {\bf 7}, 221 (2019), 1902.00134.

\bibitem{Cline:2021iff}
J.~M. Cline {\em et~al.},
\newblock Phys. Rev. D {\bf 103}, 123529 (2021), 2102.12490.

\bibitem{Chien:2015xha}
Y.~T. Chien, V.~Cirigliano, W.~Dekens, J.~de~Vries, and E.~Mereghetti,
\newblock JHEP {\bf 02}, 011 (2016), 1510.00725.

\bibitem{Bahl:2020wee}
H.~Bahl {\em et~al.},
\newblock JHEP {\bf 11}, 127 (2020), 2007.08542.

\bibitem{Engel:2013lsa}
J.~Engel, M.~J. Ramsey-Musolf, and U.~van Kolck,
\newblock Prog. Part. Nucl. Phys. {\bf 71}, 21 (2013), 1303.2371.

\bibitem{ACME:2018yjb}
ACME, V.~Andreev {\em et~al.},
\newblock Nature {\bf 562}, 355 (2018).

\bibitem{Abel:2020pzs}
C.~Abel {\em et~al.},
\newblock Phys. Rev. Lett. {\bf 124}, 081803 (2020), 2001.11966.

\bibitem{PhysRevC.97.012501}
T.~M. Ito {\em et~al.},
\newblock Phys. Rev. C {\bf 97}, 012501 (2018).

\bibitem{nEDM:2019qgk}
nEDM, M.~W. Ahmed {\em et~al.},
\newblock JINST {\bf 14}, P11017 (2019), 1908.09937.

\bibitem{Wurm:2019yfj}
D.~Wurm {\em et~al.},
\newblock EPJ Web Conf. {\bf 219}, 02006 (2019), 1911.09161.

\bibitem{n2EDM:2021yah}
n2EDM, N.~J. Ayres {\em et~al.},
\newblock Eur. Phys. J. C {\bf 81}, 512 (2021), 2101.08730.

\bibitem{2013NJPh...15e3034T}
M.~R. {Tarbutt}, B.~E. {Sauer}, J.~J. {Hudson}, and E.~A. {Hinds},
\newblock New Journal of Physics {\bf 15}, 053034 (2013), 1302.2870.

\bibitem{PhysRevLett.119.133002}
I.~Kozyryev and N.~R. Hutzler,
\newblock Phys. Rev. Lett. {\bf 119}, 133002 (2017).

\bibitem{2018EPJD...72..197A}
P.~{Aggarwal} {\em et~al.},
\newblock European Physical Journal D {\bf 72}, 197 (2018), 1804.10012.

\bibitem{Riotto:1995hh}
A.~Riotto,
\newblock Phys. Rev. {\bf D53}, 5834 (1996), hep-ph/9510271.

\bibitem{Joyce:1994zt}
M.~Joyce, T.~Prokopec, and N.~Turok,
\newblock Phys. Rev. {\bf D53}, 2958 (1996), hep-ph/9410282.

\bibitem{Cline:2020jre}
J.~M. Cline and K.~Kainulainen,
\newblock Phys. Rev. D {\bf 101}, 063525 (2020), 2001.00568.

\bibitem{Cline:2021dkf}
J.~M. Cline and B.~Laurent,
\newblock Phys. Rev. D {\bf 104}, 083507 (2021), 2108.04249.

\bibitem{Postma:2021zux}
M.~Postma,
\newblock JHEP {\bf 09}, 055 (2021), 2107.05971.

\bibitem{Kainulainen:2021oqs}
K.~Kainulainen,
\newblock JCAP {\bf 11}, 042 (2021), 2108.08336.

\bibitem{Dorsch:2021ubz}
G.~C. Dorsch, S.~J. Huber, and T.~Konstandin,
\newblock JCAP {\bf 08}, 020 (2021), 2106.06547.

\bibitem{Moore:1995si}
G.~D. Moore and T.~Prokopec,
\newblock Phys. Rev. D {\bf 52}, 7182 (1995), hep-ph/9506475.

\bibitem{Laurent:2020gpg}
B.~Laurent and J.~M. Cline,
\newblock Phys. Rev. D {\bf 102}, 063516 (2020), 2007.10935.

\bibitem{Dorsch:2021nje}
G.~C. Dorsch, S.~J. Huber, and T.~Konstandin,
\newblock (2021), 2112.12548.

\bibitem{Konstandin:2010dm}
T.~Konstandin and J.~M. No,
\newblock JCAP {\bf 02}, 008 (2011), 1011.3735.

\bibitem{BarrosoMancha:2020fay}
M.~Barroso~Mancha, T.~Prokopec, and B.~Swiezewska,
\newblock JHEP {\bf 01}, 070 (2021), 2005.10875.

\bibitem{Balaji:2020yrx}
S.~Balaji, M.~Spannowsky, and C.~Tamarit,
\newblock JCAP {\bf 03}, 051 (2021), 2010.08013.

\bibitem{Ai:2021kak}
W.-Y. Ai, B.~Garbrecht, and C.~Tamarit,
\newblock (2021), 2109.13710.

\bibitem{Ahriche:2007jp}
A.~Ahriche,
\newblock Phys. Rev. D {\bf 75}, 083522 (2007), hep-ph/0701192.

\bibitem{Funakubo:2009eg}
K.~Funakubo and E.~Senaha,
\newblock Phys. Rev. D {\bf 79}, 115024 (2009), 0905.2022.

\bibitem{Fuyuto:2014yia}
K.~Fuyuto and E.~Senaha,
\newblock Phys. Rev. D {\bf 90}, 015015 (2014), 1406.0433.

\bibitem{Fuyuto:2015jha}
K.~Fuyuto and E.~Senaha,
\newblock Phys. Lett. B {\bf 747}, 152 (2015), 1504.04291.

\bibitem{Croon:2020cgk}
D.~Croon, O.~Gould, P.~Schicho, T.~V.~I. Tenkanen, and G.~White,
\newblock JHEP {\bf 04}, 055 (2021), 2009.10080.

\bibitem{Gould:2021oba}
O.~Gould and T.~V.~I. Tenkanen,
\newblock JHEP {\bf 06}, 069 (2021), 2104.04399.

\bibitem{Caprini:2019egz}
C.~Caprini {\em et~al.},
\newblock JCAP {\bf 03}, 024 (2020), 1910.13125.

\bibitem{Lewicki:2021pgr}
M.~Lewicki, M.~Merchand, and M.~Zych,
\newblock (2021), 2111.02393.

\bibitem{Kawamura:2020pcg}
S.~Kawamura {\em et~al.},
\newblock PTEP {\bf 2021}, 05A105 (2021), 2006.13545.

\bibitem{Harry:2006fi}
G.~M. Harry, P.~Fritschel, D.~A. Shaddock, W.~Folkner, and E.~S. Phinney,
\newblock Class. Quant. Grav. {\bf 23}, 4887 (2006),
\newblock [Erratum: Class.Quant.Grav. 23, 7361 (2006)].

\bibitem{Elor:2018twp}
G.~Elor, M.~Escudero, and A.~Nelson,
\newblock Phys. Rev. D {\bf 99}, 035031 (2019), 1810.00880.

\bibitem{Elor:2020tkc}
G.~Elor and R.~McGehee,
\newblock Phys. Rev. D {\bf 103}, 035005 (2021), 2011.06115.

\bibitem{Elahi:2021jia}
F.~Elahi, G.~Elor, and R.~McGehee,
\newblock (2021), 2109.09751.

\bibitem{Alonso-Alvarez:2021oaj}
G.~Alonso-\'Alvarez {\em et~al.},
\newblock (2021), 2111.12712.

\bibitem{Goudzovski:2022vbt}
E.~Goudzovski {\em et~al.},
\newblock (2022), 2201.07805.

\bibitem{Belle:2021gmc}
Belle, C.~Hadjivasiliou {\em et~al.},
\newblock (2021), 2110.14086.

\bibitem{Rodriguez:2021urv}
A.~B. Rodr\'\i{}guez {\em et~al.},
\newblock Eur. Phys. J. C {\bf 81}, 964 (2021), 2106.12870.

\bibitem{Borsato:2021aum}
M.~Borsato {\em et~al.},
\newblock (2021), 2105.12668.

\bibitem{BESIII:2021slv}
BESIII, M.~Ablikim {\em et~al.},
\newblock (2021), 2110.06759.

\bibitem{Alonso-Alvarez:2021qfd}
G.~Alonso-\'Alvarez, G.~Elor, and M.~Escudero,
\newblock Phys. Rev. D {\bf 104}, 035028 (2021), 2101.02706.

\bibitem{Alonso-Alvarez:2019fym}
G.~Alonso-\'Alvarez, G.~Elor, A.~E. Nelson, and H.~Xiao,
\newblock JHEP {\bf 03}, 046 (2020), 1907.10612.

\bibitem{Choi:2009ym}
H.-M. Choi and C.-R. Ji,
\newblock Phys. Rev. D {\bf 80}, 114003 (2009), 0909.5028.

\bibitem{Gouz:2002kk}
I.~P. Gouz, V.~V. Kiselev, A.~K. Likhoded, V.~I. Romanovsky, and O.~P.
  Yushchenko,
\newblock Phys. Atom. Nucl. {\bf 67}, 1559 (2004), hep-ph/0211432.

\bibitem{FCC:2018byv}
FCC, A.~Abada {\em et~al.},
\newblock Eur. Phys. J. C {\bf 79}, 474 (2019).

\bibitem{Beneke:2005vv}
M.~Beneke and S.~Jager,
\newblock Nucl. Phys. B {\bf 751}, 160 (2006), hep-ph/0512351.

\bibitem{Hayano:1982wu}
R.~S. Hayano {\em et~al.},
\newblock Phys. Rev. Lett. {\bf 49}, 1305 (1982).

\bibitem{E949:2014gsn}
E949, A.~V. Artamonov {\em et~al.},
\newblock Phys. Rev. D {\bf 91}, 052001 (2015), 1411.3963,
\newblock [Erratum: Phys.Rev.D 91, 059903 (2015)].

\bibitem{PIENU:2017wbj}
PIENU, A.~Aguilar-Arevalo {\em et~al.},
\newblock Phys. Rev. D {\bf 97}, 072012 (2018), 1712.03275.

\bibitem{NA62:2017qcd}
NA62, E.~Cortina~Gil {\em et~al.},
\newblock Phys. Lett. B {\bf 778}, 137 (2018), 1712.00297.

\bibitem{PIENU:2019usb}
PIENU, A.~Aguilar-Arevalo {\em et~al.},
\newblock Phys. Lett. B {\bf 798}, 134980 (2019), 1904.03269.

\bibitem{NA62:2021bji}
NA62, E.~Cortina~Gil {\em et~al.},
\newblock Phys. Lett. B {\bf 816}, 136259 (2021), 2101.12304.

\bibitem{Minkowski:1977sc}
P.~Minkowski,
\newblock Phys. Lett. {\bf 67B}, 421 (1977).

\bibitem{Mohapatra:1979ia}
R.~N. Mohapatra and G.~Senjanovic,
\newblock Phys. Rev. Lett. {\bf 44}, 912 (1980).

\bibitem{Glashow:1979nm}
S.~Glashow,
\newblock NATO Sci. Ser. B {\bf 61}, 687 (1980).

\bibitem{GellMann:1980vs}
M.~Gell-Mann, P.~Ramond, and R.~Slansky,
\newblock Conf. Proc. {\bf C790927}, 315 (1979), 1306.4669.

\bibitem{Yanagida:1980xy}
T.~Yanagida,
\newblock Prog. Theor. Phys. {\bf 64}, 1103 (1980).

\bibitem{Schechter:1980gr}
J.~Schechter and J.~W.~F. Valle,
\newblock Phys. Rev. {\bf D22}, 2227 (1980).

\bibitem{Magg:1980ut}
M.~Magg and C.~Wetterich,
\newblock Phys. Lett. B {\bf 94}, 61 (1980).

\bibitem{Cheng:1980qt}
T.~P. Cheng and L.-F. Li,
\newblock Phys. Rev. D {\bf 22}, 2860 (1980).

\bibitem{Lazarides:1980nt}
G.~Lazarides, Q.~Shafi, and C.~Wetterich,
\newblock Nucl. Phys. B {\bf 181}, 287 (1981).

\bibitem{Mohapatra:1980yp}
R.~N. Mohapatra and G.~Senjanovic,
\newblock Phys. Rev. D {\bf 23}, 165 (1981).

\bibitem{Foot:1988aq}
R.~Foot, H.~Lew, X.~G. He, and G.~C. Joshi,
\newblock Z. Phys. C {\bf 44}, 441 (1989).

\bibitem{Cai:2017jrq}
Y.~Cai, J.~Herrero-Garc\'\i{}a, M.~A. Schmidt, A.~Vicente, and R.~R. Volkas,
\newblock Front. in Phys. {\bf 5}, 63 (2017), 1706.08524.

\bibitem{Restrepo:2013aga}
D.~Restrepo, O.~Zapata, and C.~E. Yaguna,
\newblock JHEP {\bf 11}, 011 (2013), 1308.3655.

\bibitem{Fukugita:1986hr}
M.~Fukugita and T.~Yanagida,
\newblock Phys. Lett. {\bf B174}, 45 (1986).

\bibitem{Canetti:2012zc}
L.~Canetti, M.~Drewes, and M.~Shaposhnikov,
\newblock New J. Phys. {\bf 14}, 095012 (2012), 1204.4186.

\bibitem{Ma:1998dx}
E.~Ma and U.~Sarkar,
\newblock Phys. Rev. Lett. {\bf 80}, 5716 (1998), hep-ph/9802445.

\bibitem{Hambye:2003rt}
T.~Hambye, Y.~Lin, A.~Notari, M.~Papucci, and A.~Strumia,
\newblock Nucl. Phys. B {\bf 695}, 169 (2004), hep-ph/0312203.

\bibitem{Albright:2003xb}
C.~H. Albright and S.~M. Barr,
\newblock Phys. Rev. D {\bf 69}, 073010 (2004), hep-ph/0312224.

\bibitem{Antusch:2004xy}
S.~Antusch and S.~F. King,
\newblock Phys. Lett. B {\bf 597}, 199 (2004), hep-ph/0405093.

\bibitem{Hambye:2012fh}
T.~Hambye,
\newblock New J. Phys. {\bf 14}, 125014 (2012), 1212.2888.

\bibitem{DOnofrio:2014rug}
M.~D'Onofrio, K.~Rummukainen, and A.~Tranberg,
\newblock Phys. Rev. Lett. {\bf 113}, 141602 (2014), 1404.3565.

\bibitem{Davidson:2002qv}
S.~Davidson and A.~Ibarra,
\newblock Phys. Lett. B {\bf 535}, 25 (2002), hep-ph/0202239.

\bibitem{Buchmuller:2005eh}
W.~Buchmuller, R.~D. Peccei, and T.~Yanagida,
\newblock Ann. Rev. Nucl. Part. Sci. {\bf 55}, 311 (2005), hep-ph/0502169.

\bibitem{Davidson:2008bu}
S.~Davidson, E.~Nardi, and Y.~Nir,
\newblock Phys. Rept. {\bf 466}, 105 (2008), 0802.2962.

\bibitem{Bodeker:2020ghk}
D.~Bodeker and W.~Buchmuller,
\newblock Rev. Mod. Phys. {\bf 93}, 035004 (2021), 2009.07294.

\bibitem{DiBari:2021fhs}
P.~Di~Bari,
\newblock Prog. Part. Nucl. Phys. {\bf 122}, 103913 (2022), 2107.13750.

\bibitem{Dev:2017trv}
P.~S.~B. Dev {\em et~al.},
\newblock Int. J. Mod. Phys. A {\bf 33}, 1842001 (2018), 1711.02861.

\bibitem{Drewes:2017zyw}
M.~Drewes {\em et~al.},
\newblock Int. J. Mod. Phys. A {\bf 33}, 1842002 (2018), 1711.02862.

\bibitem{Dev:2017wwc}
B.~Dev, M.~Garny, J.~Klaric, P.~Millington, and D.~Teresi,
\newblock Int. J. Mod. Phys. A {\bf 33}, 1842003 (2018), 1711.02863.

\bibitem{Biondini:2017rpb}
S.~Biondini {\em et~al.},
\newblock Int. J. Mod. Phys. A {\bf 33}, 1842004 (2018), 1711.02864.

\bibitem{Chun:2017spz}
E.~J. Chun {\em et~al.},
\newblock Int. J. Mod. Phys. A {\bf 33}, 1842005 (2018), 1711.02865.

\bibitem{Hagedorn:2017wjy}
C.~Hagedorn, R.~N. Mohapatra, E.~Molinaro, C.~C. Nishi, and S.~T. Petcov,
\newblock Int. J. Mod. Phys. A {\bf 33}, 1842006 (2018), 1711.02866.

\bibitem{Deppisch:2013jxa}
F.~F. Deppisch, J.~Harz, and M.~Hirsch,
\newblock Phys. Rev. Lett. {\bf 112}, 221601 (2014), 1312.4447.

\bibitem{Deppisch:2015yqa}
F.~F. Deppisch, J.~Harz, M.~Hirsch, W.-C. Huang, and H.~P\"as,
\newblock Phys. Rev. D {\bf 92}, 036005 (2015), 1503.04825.

\bibitem{Akhmedov:1998qx}
E.~K. Akhmedov, V.~A. Rubakov, and A.~Y. Smirnov,
\newblock Phys. Rev. Lett. {\bf 81}, 1359 (1998), hep-ph/9803255.

\bibitem{Asaka:2005pn}
T.~Asaka and M.~Shaposhnikov,
\newblock Phys. Lett. B {\bf 620}, 17 (2005), hep-ph/0505013.

\bibitem{Pilaftsis:2003gt}
A.~Pilaftsis and T.~E.~J. Underwood,
\newblock Nucl. Phys. B {\bf 692}, 303 (2004), hep-ph/0309342.

\bibitem{Pilaftsis:2005rv}
A.~Pilaftsis and T.~E.~J. Underwood,
\newblock Phys. Rev. D {\bf 72}, 113001 (2005), hep-ph/0506107.

\bibitem{Dev:2014pfm}
P.~S.~B. Dev, P.~Millington, A.~Pilaftsis, and D.~Teresi,
\newblock Nucl. Phys. B {\bf 886}, 569 (2014), 1404.1003.

\bibitem{Agrawal:2021dbo}
P.~Agrawal {\em et~al.},
\newblock Eur. Phys. J. C {\bf 81}, 1015 (2021), 2102.12143.

\bibitem{Atre:2009rg}
A.~Atre, T.~Han, S.~Pascoli, and B.~Zhang,
\newblock JHEP {\bf 05}, 030 (2009), 0901.3589.

\bibitem{Deppisch:2015qwa}
F.~F. Deppisch, P.~S.~B. Dev, and A.~Pilaftsis,
\newblock New J. Phys. {\bf 17}, 075019 (2015), 1502.06541.

\bibitem{Cai:2017mow}
Y.~Cai, T.~Han, T.~Li, and R.~Ruiz,
\newblock Front. in Phys. {\bf 6}, 40 (2018), 1711.02180.

\bibitem{Liu:1993tg}
J.~Liu and G.~Segre,
\newblock Phys. Rev. D {\bf 48}, 4609 (1993), hep-ph/9304241.

\bibitem{Flanz:1994yx}
M.~Flanz, E.~A. Paschos, and U.~Sarkar,
\newblock Phys. Lett. B {\bf 345}, 248 (1995), hep-ph/9411366,
\newblock [Erratum: Phys.Lett.B 384, 487--487 (1996), Erratum: Phys.Lett.B 382,
  447--447 (1996)].

\bibitem{Flanz:1996fb}
M.~Flanz, E.~A. Paschos, U.~Sarkar, and J.~Weiss,
\newblock Phys. Lett. B {\bf 389}, 693 (1996), hep-ph/9607310.

\bibitem{Covi:1996wh}
L.~Covi, E.~Roulet, and F.~Vissani,
\newblock Phys. Lett. B {\bf 384}, 169 (1996), hep-ph/9605319.

\bibitem{Covi:1996fm}
L.~Covi and E.~Roulet,
\newblock Phys. Lett. B {\bf 399}, 113 (1997), hep-ph/9611425.

\bibitem{Pilaftsis:1997jf}
A.~Pilaftsis,
\newblock Phys. Rev. D {\bf 56}, 5431 (1997), hep-ph/9707235.

\bibitem{Buchmuller:1997yu}
W.~Buchmuller and M.~Plumacher,
\newblock Phys. Lett. B {\bf 431}, 354 (1998), hep-ph/9710460.

\bibitem{Klaric:2020phc}
J.~Klari\'c, M.~Shaposhnikov, and I.~Timiryasov,
\newblock Phys. Rev. Lett. {\bf 127}, 111802 (2021), 2008.13771.

\bibitem{Klaric:2021cpi}
J.~Klari\'c, M.~Shaposhnikov, and I.~Timiryasov,
\newblock Phys. Rev. D {\bf 104}, 055010 (2021), 2103.16545.

\bibitem{Drewes:2021nqr}
M.~Drewes, Y.~Georis, and J.~Klari\'c,
\newblock Phys. Rev. Lett. {\bf 128}, 051801 (2022), 2106.16226.

\bibitem{Garbrecht:2018mrp}
B.~Garbrecht,
\newblock Prog. Part. Nucl. Phys. {\bf 110}, 103727 (2020), 1812.02651.

\bibitem{Ghiglieri:2018wbs}
J.~Ghiglieri and M.~Laine,
\newblock JHEP {\bf 02}, 014 (2019), 1811.01971.

\bibitem{Ghiglieri:2020ulj}
J.~Ghiglieri and M.~Laine,
\newblock JCAP {\bf 07}, 012 (2020), 2004.10766.

\bibitem{Canetti:2010aw}
L.~Canetti and M.~Shaposhnikov,
\newblock JCAP {\bf 09}, 001 (2010), 1006.0133.

\bibitem{Canetti:2012vf}
L.~Canetti, M.~Drewes, and M.~Shaposhnikov,
\newblock Phys. Rev. Lett. {\bf 110}, 061801 (2013), 1204.3902.

\bibitem{Canetti:2012kh}
L.~Canetti, M.~Drewes, T.~Frossard, and M.~Shaposhnikov,
\newblock Phys. Rev. D {\bf 87}, 093006 (2013), 1208.4607.

\bibitem{Dev:2014oar}
P.~S.~B. Dev, P.~Millington, A.~Pilaftsis, and D.~Teresi,
\newblock Nucl. Phys. B {\bf 891}, 128 (2015), 1410.6434.

\bibitem{Hernandez:2015wna}
P.~Hern\'andez, M.~Kekic, J.~L\'opez-Pav\'on, J.~Racker, and N.~Rius,
\newblock JHEP {\bf 10}, 067 (2015), 1508.03676.

\bibitem{Shuve:2014zua}
B.~Shuve and I.~Yavin,
\newblock Phys. Rev. D {\bf 89}, 075014 (2014), 1401.2459.

\bibitem{Drewes:2016gmt}
M.~Drewes, B.~Garbrecht, D.~Gueter, and J.~Klaric,
\newblock JHEP {\bf 12}, 150 (2016), 1606.06690.

\bibitem{Hernandez:2016kel}
P.~Hern\'andez, M.~Kekic, J.~L\'opez-Pav\'on, J.~Racker, and J.~Salvado,
\newblock JHEP {\bf 08}, 157 (2016), 1606.06719.

\bibitem{Drewes:2016jae}
M.~Drewes, B.~Garbrecht, D.~Gueter, and J.~Klaric,
\newblock JHEP {\bf 08}, 018 (2017), 1609.09069.

\bibitem{Eijima:2018qke}
S.~Eijima, M.~Shaposhnikov, and I.~Timiryasov,
\newblock JHEP {\bf 07}, 077 (2019), 1808.10833.

\bibitem{Boiarska:2019jcw}
I.~Boiarska {\em et~al.},
\newblock (2019), 1902.04535.

\bibitem{Verhaaren:2022ify}
C.~B. Verhaaren {\em et~al.},
\newblock {Searches for Long-Lived Particles at the Future FCC-ee},
\newblock in {\em {2022 Snowmass Summer Study}}, 2022, 2203.05502.

\bibitem{Caputo:2017pit}
A.~Caputo, P.~Hernandez, J.~Lopez-Pavon, and J.~Salvado,
\newblock JHEP {\bf 06}, 112 (2017), 1704.08721.

\bibitem{Drewes:2018gkc}
M.~Drewes, J.~Hajer, J.~Klaric, and G.~Lanfranchi,
\newblock JHEP {\bf 07}, 105 (2018), 1801.04207.

\bibitem{Esteban:2020cvm}
I.~Esteban, M.~C. Gonzalez-Garcia, M.~Maltoni, T.~Schwetz, and A.~Zhou,
\newblock JHEP {\bf 09}, 178 (2020), 2007.14792.

\bibitem{DUNE:2020jqi}
DUNE, B.~Abi {\em et~al.},
\newblock Eur. Phys. J. C {\bf 80}, 978 (2020), 2006.16043.

\bibitem{DUNE:2020ypp}
DUNE, B.~Abi {\em et~al.},
\newblock (2020), 2002.03005.

\bibitem{Antusch:2017ebe}
S.~Antusch, E.~Cazzato, and O.~Fischer,
\newblock Mod. Phys. Lett. A {\bf 34}, 1950061 (2019), 1709.03797.

\bibitem{Antusch:2017pkq}
S.~Antusch {\em et~al.},
\newblock JHEP {\bf 09}, 124 (2018), 1710.03744.

\bibitem{Asaka:2016zib}
T.~Asaka, S.~Eijima, and H.~Ishida,
\newblock Phys. Lett. B {\bf 762}, 371 (2016), 1606.06686.

\bibitem{Drewes:2016lqo}
M.~Drewes and S.~Eijima,
\newblock Phys. Lett. B {\bf 763}, 72 (2016), 1606.06221.

\bibitem{Drewes:2013gca}
M.~Drewes,
\newblock Int. J. Mod. Phys. E {\bf 22}, 1330019 (2013), 1303.6912.

\bibitem{Antusch:2014woa}
S.~Antusch and O.~Fischer,
\newblock JHEP {\bf 10}, 094 (2014), 1407.6607.

\bibitem{Gorbunov:2014ypa}
D.~Gorbunov and I.~Timiryasov,
\newblock Phys. Lett. B {\bf 745}, 29 (2015), 1412.7751.

\bibitem{Drewes:2015iva}
M.~Drewes and B.~Garbrecht,
\newblock Nucl. Phys. B {\bf 921}, 250 (2017), 1502.00477.

\bibitem{deGouvea:2015euy}
A.~de~Gouv\^ea and A.~Kobach,
\newblock Phys. Rev. D {\bf 93}, 033005 (2016), 1511.00683.

\bibitem{Fernandez-Martinez:2016lgt}
E.~Fernandez-Martinez, J.~Hernandez-Garcia, and J.~Lopez-Pavon,
\newblock JHEP {\bf 08}, 033 (2016), 1605.08774.

\bibitem{Chrzaszcz:2019inj}
M.~Chrzaszcz {\em et~al.},
\newblock Eur. Phys. J. C {\bf 80}, 569 (2020), 1908.02302.

\bibitem{Bolton:2019pcu}
P.~D. Bolton, F.~F. Deppisch, and P.~S.~B. Dev,
\newblock JHEP {\bf 03}, 170 (2020), 1912.03058.

\bibitem{Shaposhnikov:2006nn}
M.~Shaposhnikov,
\newblock Nucl. Phys. B {\bf 763}, 49 (2007), hep-ph/0605047.

\bibitem{Kersten:2007vk}
J.~Kersten and A.~Y. Smirnov,
\newblock Phys. Rev. D {\bf 76}, 073005 (2007), 0705.3221.

\bibitem{Moffat:2017feq}
K.~Moffat, S.~Pascoli, and C.~Weiland,
\newblock (2017), 1712.07611.

\bibitem{Ibarra:2010xw}
A.~Ibarra, E.~Molinaro, and S.~T. Petcov,
\newblock JHEP {\bf 09}, 108 (2010), 1007.2378.

\bibitem{Anamiati:2016uxp}
G.~Anamiati, M.~Hirsch, and E.~Nardi,
\newblock JHEP {\bf 10}, 010 (2016), 1607.05641.

\bibitem{Boyanovsky:2014una}
D.~Boyanovsky,
\newblock Phys. Rev. D {\bf 90}, 105024 (2014), 1409.4265.

\bibitem{Antusch:2020pnn}
S.~Antusch and J.~Rosskopp,
\newblock JHEP {\bf 03}, 170 (2021), 2012.05763.

\bibitem{Kayser:1982br}
B.~Kayser,
\newblock Phys. Rev. D {\bf 26}, 1662 (1982).

\bibitem{Balantekin:2018ukw}
A.~B. Balantekin, A.~de~Gouv\^ea, and B.~Kayser,
\newblock Phys. Lett. B {\bf 789}, 488 (2019), 1808.10518.

\bibitem{Tastet:2019nqj}
J.-L. Tastet and I.~Timiryasov,
\newblock JHEP {\bf 04}, 005 (2020), 1912.05520.

\bibitem{Blondel:2021mss}
A.~Blondel, A.~de~Gouv\^ea, and B.~Kayser,
\newblock Phys. Rev. D {\bf 104}, 055027 (2021), 2105.06576.

\bibitem{deGouvea:2021rpa}
A.~de~Gouv\^ea, P.~J. Fox, B.~J. Kayser, and K.~J. Kelly,
\newblock Phys. Rev. D {\bf 105}, 015019 (2022), 2109.10358.

\bibitem{Dib:2016wge}
C.~O. Dib, C.~S. Kim, K.~Wang, and J.~Zhang,
\newblock Phys. Rev. D {\bf 94}, 013005 (2016), 1605.01123.

\bibitem{Dib:2017iva}
C.~O. Dib, C.~S. Kim, and K.~Wang,
\newblock Phys. Rev. D {\bf 95}, 115020 (2017), 1703.01934.

\bibitem{Das:2017hmg}
A.~Das, P.~S.~B. Dev, and R.~N. Mohapatra,
\newblock Phys. Rev. D {\bf 97}, 015018 (2018), 1709.06553.

\bibitem{Hernandez:2018cgc}
P.~Hern\'andez, J.~Jones-P\'erez, and O.~Suarez-Navarro,
\newblock Eur. Phys. J. C {\bf 79}, 220 (2019), 1810.07210.

\bibitem{Abada:2019bac}
A.~Abada, C.~Hati, X.~Marcano, and A.~M. Teixeira,
\newblock JHEP {\bf 09}, 017 (2019), 1904.05367.

\bibitem{Drewes:2019byd}
M.~Drewes, J.~Klari\'c, and P.~Klose,
\newblock JHEP {\bf 11}, 032 (2019), 1907.13034.

\bibitem{Caputo:2016ojx}
A.~Caputo, P.~Hernandez, M.~Kekic, J.~L\'opez-Pav\'on, and J.~Salvado,
\newblock Eur. Phys. J. C {\bf 77}, 258 (2017), 1611.05000.

\bibitem{Cvetic:2015naa}
G.~Cvetic, C.~Dib, C.~S. Kim, and J.~Zamora-Saa,
\newblock Symmetry {\bf 7}, 726 (2015), 1503.01358.

\bibitem{Cvetic:2015ura}
G.~Cvetic, C.~S. Kim, R.~Kogerler, and J.~Zamora-Saa,
\newblock Phys. Rev. D {\bf 92}, 013015 (2015), 1505.04749.

\bibitem{Cvetic:2018elt}
G.~Cveti\v{c}, A.~Das, and J.~Zamora-Sa\'a,
\newblock J. Phys. G {\bf 46}, 075002 (2019), 1805.00070.

\bibitem{Cvetic:2019rms}
G.~Cveti\v{c}, A.~Das, S.~Tapia, and J.~Zamora-Sa\'a,
\newblock J. Phys. G {\bf 47}, 015001 (2020), 1905.03097.

\bibitem{Asaka:2005an}
T.~Asaka, S.~Blanchet, and M.~Shaposhnikov,
\newblock Phys. Lett. B {\bf 631}, 151 (2005), hep-ph/0503065.

\bibitem{CHARM:1985nku}
CHARM, F.~Bergsma {\em et~al.},
\newblock Phys. Lett. B {\bf 166}, 473 (1986).

\bibitem{Abela:1981nf}
R.~Abela {\em et~al.},
\newblock Phys. Lett. B {\bf 105}, 263 (1981),
\newblock [Erratum: Phys.Lett.B 106, 513 (1981)].

\bibitem{Yamazaki:1984sj}
T.~Yamazaki {\em et~al.},
\newblock Conf. Proc. C {\bf 840719}, 262 (1984).

\bibitem{Bernardi:1987ek}
G.~Bernardi {\em et~al.},
\newblock Phys. Lett. B {\bf 203}, 332 (1988).

\bibitem{NuTeV:1999kej}
NuTeV, E815, A.~Vaitaitis {\em et~al.},
\newblock Phys. Rev. Lett. {\bf 83}, 4943 (1999), hep-ex/9908011.

\bibitem{Vaitaitis:2000vc}
A.~G. Vaitaitis,
\newblock {\em {Search for neutral heavy leptons in a high-energy neutrino
  beam}},
\newblock PhD thesis, Columbia U., 2000.

\bibitem{CMS:2018iaf}
CMS, A.~M. Sirunyan {\em et~al.},
\newblock Phys. Rev. Lett. {\bf 120}, 221801 (2018), 1802.02965.

\bibitem{DELPHI:1996qcc}
DELPHI, P.~Abreu {\em et~al.},
\newblock Z. Phys. C {\bf 74}, 57 (1997),
\newblock [Erratum: Z.Phys.C 75, 580 (1997)].

\bibitem{ATLAS:2019kpx}
ATLAS, G.~Aad {\em et~al.},
\newblock JHEP {\bf 10}, 265 (2019), 1905.09787.

\bibitem{CMS:2022fut}
CMS, A.~Tumasyan {\em et~al.},
\newblock (2022), 2201.05578.

\bibitem{Sabti:2020yrt}
N.~Sabti, A.~Magalich, and A.~Filimonova,
\newblock JCAP {\bf 11}, 056 (2020), 2006.07387.

\bibitem{Boyarsky:2020dzc}
A.~Boyarsky, M.~Ovchynnikov, O.~Ruchayskiy, and V.~Syvolap,
\newblock Phys. Rev. D {\bf 104}, 023517 (2021), 2008.00749.

\bibitem{Izaguirre:2015pga}
E.~Izaguirre and B.~Shuve,
\newblock Phys. Rev. D {\bf 91}, 093010 (2015), 1504.02470.

\bibitem{Drewes:2019fou}
M.~Drewes and J.~Hajer,
\newblock JHEP {\bf 02}, 070 (2020), 1903.06100.

\bibitem{Pascoli:2018heg}
S.~Pascoli, R.~Ruiz, and C.~Weiland,
\newblock JHEP {\bf 06}, 049 (2019), 1812.08750.

\bibitem{Ballett:2019bgd}
P.~Ballett, T.~Boschi, and S.~Pascoli,
\newblock JHEP {\bf 03}, 111 (2020), 1905.00284.

\bibitem{FASER:2018eoc}
FASER, A.~Ariga {\em et~al.},
\newblock Phys. Rev. D {\bf 99}, 095011 (2019), 1811.12522.

\bibitem{SHiP:2018xqw}
SHiP, C.~Ahdida {\em et~al.},
\newblock JHEP {\bf 04}, 077 (2019), 1811.00930.

\bibitem{Gorbunov:2020rjx}
D.~Gorbunov, I.~Krasnov, Y.~Kudenko, and S.~Suvorov,
\newblock Phys. Lett. B {\bf 810}, 135817 (2020), 2004.07974.

\bibitem{Curtin:2018mvb}
D.~Curtin {\em et~al.},
\newblock Rept. Prog. Phys. {\bf 82}, 116201 (2019), 1806.07396.

\bibitem{Aielli:2019ivi}
G.~Aielli {\em et~al.},
\newblock Eur. Phys. J. C {\bf 80}, 1177 (2020), 1911.00481.

\bibitem{Antusch:2016ejd}
S.~Antusch, E.~Cazzato, and O.~Fischer,
\newblock Int. J. Mod. Phys. A {\bf 32}, 1750078 (2017), 1612.02728.

\bibitem{Abdullahi:2022jlv}
A.~M. Abdullahi {\em et~al.},
\newblock {The Present and Future Status of Heavy Neutral Leptons},
\newblock 2022, 2203.08039.

\bibitem{Deppisch:2017ecm}
F.~F. Deppisch, L.~Graf, J.~Harz, and W.-C. Huang,
\newblock Phys. Rev. D {\bf 98}, 055029 (2018), 1711.10432.

\bibitem{Harz:2021psp}
J.~Harz, M.~J. Ramsey-Musolf, T.~Shen, and S.~Urrutia-Quiroga,
\newblock (2021), 2106.10838.

\bibitem{AristizabalSierra:2013lyx}
D.~Aristizabal~Sierra, C.~S. Fong, E.~Nardi, and E.~Peinado,
\newblock JCAP {\bf 02}, 013 (2014), 1309.4770.

\bibitem{Frandsen:2018jfi}
M.~T. Frandsen, C.~Hagedorn, W.-C. Huang, E.~Molinaro, and H.~P\"as,
\newblock Phys. Lett. B {\bf 782}, 387 (2018), 1801.09314.

\end{thebibliography}
